\documentclass[twocolumn, twocolappendix]{aastex631}

\let\tablenum\relax

\usepackage{amsmath}
\usepackage{fancyvrb}
\usepackage{siunitx}[=v2]
\setcitestyle{notesep={;}}
\newcommand{\lya}{\ensuremath{\text{Ly}\alpha}}

\newcommand{\arcfull}{SGAS J122651.3+215220}
\newcommand{\arc}{SGASJ1226}

\newcommand{\name}{SGASJ1226}
\DeclareSIUnit\jansky{Jy}
\DeclareSIUnit\erg{erg}
\DeclareSIUnit\parsec{pc}
\DeclareSIUnit\arcsec{arcsec}
\DeclareSIUnit\arcsect{^{\prime\prime}}
\DeclareSIUnit\msun{M_\odot}
\DeclareSIUnit\lumsol{L_\odot}
\DeclareSIUnit\zsun{Z_\odot}
\DeclareSIUnit\year{yr}
\DeclareSIUnit\beam{beam}
\DeclareSIUnit\angstrom{\text {Å}}

\shortauthors{Solimano et al.}

\begin{document}

  \title{Revealing the Nature of a \lya~Halo in a Strongly Lensed Interacting System at $z=2.92$}

  \author[0000-0001-6629-0379]{%
	  Manuel Solimano
  }
  \affiliation{%
	  N\'ucleo de Astronom\'ia, Facultad de Ingenier\'ia y Ciencias,
  Universidad Diego Portales, \\ Av. Ej\'ercito Libertador 441,  Santiago, Chile.
  }
  
  \author[0000-0003-3926-1411]{%
     Jorge Gonz\'alez-L\'opez
  }
  \affiliation{%
    Las Campanas Observatory, Carnegie Institution of Washington, %
    Casilla 601, La Serena, Chile
  }
  \affiliation{%
	  N\'ucleo de Astronom\'ia, Facultad de Ingenier\'ia y Ciencias, 
  Universidad Diego Portales, \\ Av. Ej\'ercito Libertador 441,  Santiago, Chile.
  }

  \author[0000-0002-6290-3198]{%
    Manuel Aravena
  }
  \affiliation{%
	  N\'ucleo de Astronom\'ia, Facultad de Ingenier\'ia y Ciencias, 
  Universidad Diego Portales, \\ Av. Ej\'ercito Libertador 441,  Santiago, Chile.
  }
  
  \author[0000-0002-2368-6469]{%
    Evelyn J. Johnston
  }
  \affiliation{%
	  N\'ucleo de Astronom\'ia, Facultad de Ingenier\'ia y Ciencias, 
  Universidad Diego Portales, \\ Av. Ej\'ercito Libertador 441,  Santiago, Chile.
  }
  
  \author[0000-0002-8876-267X]{%
    Crist\'obal Moya-Sierralta
  }
  \affiliation{%
    Instituto de Astrof\'isica Facultad de F\'isica, Pontificia Universidad Cat\'olica de Chile, \\ Av. Vicu\~na Mackenna 4860, 782-0436, Macul, Santiago, Chile
  }
  
  \author[0000-0003-0151-0718]{%
    Luis F. Barrientos
  }
  \affiliation{%
   Instituto de Astrof\'isica Facultad de F\'isica, Pontificia Universidad Cat\'olica de Chile, \\ Av. Vicu\~na Mackenna 4860, 782-0436, Macul, Santiago, Chile
  }
  
  \author[0000-0003-1074-4807]{%
    Matthew B. Bayliss
  }
  \affiliation{%
    Department of Physics, University of Cincinnati, Cincinnati, OH 45221, USA
  }
  
  \author[0000-0003-1370-5010]{
    Michael Gladders
  }
  \affiliation{
    Department of Astronomy \& Astrophysics, The University of Chicago, 5640 S. Ellis Ave., Chicago, IL 60637, USA  
  }
  
  \author[0000-0001-8581-932X]{%
     Leopoldo Infante
  }
  \affiliation{%
    Las Campanas Observatory, Carnegie Institution of Washington, %
    Casilla 601, La Serena, Chile
  }
  \affiliation{%
	  N\'ucleo de Astronom\'ia, Facultad de Ingenier\'ia y Ciencias, 
  Universidad Diego Portales, \\ Av. Ej\'ercito Libertador 441,  Santiago, Chile.
  }
  
  \author[0000-0002-7864-3327]{
    C\'edric Ledoux
  }
  \affiliation{
    European Southern Observatory, Alonso de C\'ordova 3107, Vitacura, Casilla 19001, Santiago, Chile
  }
  
  \author[0000-0003-0389-0902]{
    Sebasti\'an L\'opez
  }
  \affiliation{
    Departamento de Astronom\'ia, Universidad de Chile, Casilla 36-D, Santiago, Chile
  }
  
  \author[0000-0003-0536-3081]{
    Suraj Poudel  
  }
  \affiliation{
    Instituto de F\'isica, Pontificia Universidad Cat\'olica de Valpara\'iso, Casilla 4059, Valpara\'iso, Chile
  }
  
  \author[0000-0002-7627-6551]{
    Jane R. Rigby
  }
  \affiliation{
    Observational Cosmology Lab, NASA Goddard Space Flight Center, \\ Code 665, 8800 Greenbelt Rd, Greenbelt, MD 20771, USA
  }
  
  \author[0000-0002-7559-0864]{
    Keren Sharon
  }
  \affiliation{
    Department of Astronomy, University of Michigan, 1085 S. University Ave., Ann Arbor, MI 48109, USA
  }
  
  \author[0000-0002-1883-4252]{
   Nicol\'as Tejos
  }
  \affiliation{
  Instituto de F\'isica, Pontificia Universidad Cat\'olica de Valpara\'iso, Casilla 4059, Valpara\'iso, Chile
  }
  

  \begin{abstract}
    Spatially extended halos of \ion{H}{1} \lya~emission are now ubiquitously found around high-redshift star-forming galaxies. But our understanding of the nature and powering mechanisms of these halos is still hampered by the complex radiative transfer effects of the  \lya~line and limited angular resolution. In this paper, we present resolved Multi Unit Spectroscopic Explorer (MUSE) observations of \arcfull, a strongly-lensed pair of $L^{*}$ galaxies at $z=2.92$ embedded in a \lya~halo of $L_{\lya}=\SI{6.2+-1.3e42}{\erg\per\second}$. Globally, the system shows a line profile that is markedly asymmetric and redshifted, but its width and peak shift vary significantly across the halo. By fitting the spatially binned \lya~spectra with a collection of radiative transfer galactic wind models, we infer a mean outflow expansion velocity of $\approx\SI{211}{\kilo\meter\per\second}$, with higher values preferentially found on both sides of the system's major axis. The velocity of the outflow is validated with the blueshift of low-ionization metal absorption lines in the spectra of the central galaxies. We also identify a faint ($M_{1500} \approx -16.7$) companion detected in both \lya~and the continuum, whose properties are in agreement with a predicted population of satellite galaxies that contribute to the extended \lya~emission. Finally, we briefly discuss the impact of the interaction between the central galaxies on the properties of the halo and the possibility of \textit{in situ} fluorescent \lya~production.  
  \end{abstract}

  \section{Introduction}\label{sec:intro}
  
  Galaxies are embedded in envelopes of a multiphase gas known as the circumgalactic medium (CGM). The CGM exists at an intermediate scale between the interstellar medium (ISM) and the intergalactic medium (IGM), and in the case of  star forming galaxies (SFGs), it contains large reservoirs of cool ($T\sim\SI{e4}{\kelvin}$) gas that feed and regulate the star formation activity in the host galaxy (see \citealt{Tumlinson2017CircumgalacticMediumReview} for a review). Therefore, understanding the kinematics, physical conditions and spatial properties of the cool CGM is critical for answering the question of how galaxies evolve \citep[e.g.,][and references therein]{PerouxHowk2020CosmicBaryonAndMetalCycles}. Historically, our knowledge of the CGM has been inferred from the statistical properties of intervening absorption systems toward distant quasars \citep[e.g.,][]{Churchill2000MgIIAbsorbers1,Churchill2000MgIIAbsorbers2, Peroux2003SubDLAS2Properties, Prochaska2003AgeMetallicity100DLAs, Krogager2013ComprehensiveDLAstudy, Nielsen2013Magiicat1, SanchezRamirez2016EvolutionNeutralGasDLAs}, but this technique is not  well suited for single-object studies, since all the information is obtained from one or very few lines of sight per galaxy. It is thus desirable to use spatially resolved spectroscopy of CGM emission lines to obtain a more complete picture of the gas properties in individual galaxies. Unfortunately, at $z>0.3$ the \ion{H}{1} \SI{21}{\centi\meter} line, the canonical tracer of neutral gas, becomes too faint for individual detections.  However,  at $z>2.3$ the  \ion{H}{1} \lya~transition at $\lambda_\text{rest}=\SI{1215.67}{\angstrom}$ is shifted to visible wavelengths and has emerged as a powerful alternative for investigating the neutral phase of the CGM \citep[e.g.,][]{Wisotzki2016ExtendedLyaHalosMUSE}. Besides its large intrinsic brightness, \lya~has gathered interest due to the discovery of diffuse emission extending beyond the stellar component of galaxies in deep narrowband (NB) imaging surveys \citep{Steidel2000LyaProtoCluster, Steidel2011LyaHalosGeneric, Cantalupo2012FluorescentLyaQuasarCGM, Matsuda2012DiffuseLyaHaloesDM}, which suggest that \lya~does indeed trace the CGM. These regions of extended \lya~emission around SFGs are also known in the literature as \lya~halos \citep[LAHs; ][]{Hayashino2004LargeScaleLya, Steidel2011LyaHalosGeneric, Wisotzki2016ExtendedLyaHalosMUSE, Leclerq2017MuseHudfLyaHaloes, Chen2021KbssKcwiiSurveyAzimuthalAngle}. With typical exponential scale lengths of $1-\SI{10}{\kilo\parsec}$ and luminosities $L_{\lya} \lesssim \SI{e43}{\erg\per\second}$ \citep{Ouchi2020LymanAlphaReview}, LAHs should be distinguished from the larger and more luminous (but less abundant) \lya~``blobs''  \citep[e.g.,][]{Steidel2000LyaProtoCluster, Matsuda2004SubaruLyaBlobs, Ouchi2009EoRLyaBlob, Borisova2016MUSELyaBlobsAroundQSOs, Shibuya2018SilverRushII, Drake2020IonizedAndCoolGasLyaBlob} that are typically associated with overdense regions containing several galaxies.

  Due to its resonant nature, the \lya~line becomes optically thick at very low densities ($N_\text{H} > \SI{e13}{\per\centi\meter\squared}$; \citealt{Ouchi2020LymanAlphaReview}), and thus in most environments it experiences complex radiative transfer effects that conceal the kinematics, column density, and ionization state of gas. 
  As a consequence, researchers have attempted to infer the gas properties indirectly by combining spatially resolved spectroscopy (e.g., long-slit and first-generation integral-field unit (IFU) spectrographs) with models and simulations. However, this has been preferentially done on the brightest and most extended systems (i.e., \lya~blobs with $L_{\lya}>\SI{e43}{\erg\per\second}$), as fainter halos are more challenging to detect due to sensitivity limitations. In recent years, these limitations have been relaxed thanks to the availability of high-throughput integral-field spectrographs such as the Multi Unit Spectroscopic Explorer \citep[MUSE; ][]{Bacon2010MUSEpaper} on the Very Large Telescope (VLT)  and the Keck Cosmic Web Imager \citep[KCWI; ][]{Morrissey2018KCWI} at the Keck II telescope. With these instruments, galaxy-scale LAHs are now routinely detected and they are confirmed to be ubiquitous among high redshift SFGs \citep{Wisotzki2016ExtendedLyaHalosMUSE}.
  
  Theoretical models and simulations show that LAHs form by several mechanisms. On one hand, \lya~photons produced in central \ion{H}{2} regions can propagate through the neutral gas out to the CGM in a series of resonant scattering events. On the other hand, \lya~photons in the CGM can also be produced \textit{in situ}, by means of free-bound collisional interactions, photoionization by internal or external sources, or star formation in satellite galaxies. Observationally, determining which mechanisms are at play is challenging, since the spectral shape of the line can be very similar for different underlying scenarios. Fortunately, simulations also predict LAHs to have a rich spatial substructure, featuring filaments, clumps, and satellites, as well as spatial variations in the spectral properties of the line \citep{Mitchell2018SimulatedCGM, Behrens2019LyaSimulatedGalaxiesEoR, Smith2019PhysicsOfLyaEscpeHighZ}. For this reason, high angular resolution spectroscopy of LAHs is very valuable, since it can help us understand the physical nature of LAHs and potentially disentangle some of the degeneracies that affect the interpretation of the \lya~line.
  
  In this context, a growing number of studies have exploited the power of strong gravitational lensing to resolve LAHs in a great level of detail  \citep[e.g.,][]{Swinbank2007ResolvedSpectroscopyLensedLBG, Karman2015MuseFF1, Caminha2016FaintLensedLyaBlob, Patricio2016LensedLyaHaloMUSE, Smit2017GravBoostedMuseEmissionLines,  Erb2018LyaKinematicsLensed, Claeyssens2019LyaSpectralVariations, Chen2021LensedLyaSuperWinds, Claeyssens2022LlamasPaper1}. This approach was pioneered by \citet{Swinbank2007ResolvedSpectroscopyLensedLBG}, who obtained early IFU data of a giant lensed arc corresponding to a galaxy at $z=4.8$. With MUSE, the efficiency of this technique was enhanced, providing the first tentative evidence of spatial variations in the line profile of a few lensed LAHs \citep{Smit2017GravBoostedMuseEmissionLines, Erb2018LyaKinematicsLensed}. Later on, \citet{Claeyssens2019LyaSpectralVariations} showed two examples of bright lensed LAHs with a robust measurement of the \lya~variation across the halo, finding in both cases a broader and redder line toward the outskirts of the halo. Although similar results have been  obtained in a handful of nonlensed halos in the MUSE \textit{Hubble} Ultra Deep Field \citep[hereafter UDF;][]{Leclercq2020SpatiallyResolvedLyaHalosMHUDF}, the spatial scales reached with lensing remain unrivaled. 
  
  In this paper we present MUSE observations of \object{\arcfull}~(hereafter \name), a lensed, multiply imaged pair of SFGs at $z=2.92$. The main arc was discovered as a \textit{u}-band dropout \citep{Koester2010ArcJ1226Discovery} and thanks to its high apparent brightness ($r=20.6$ mag), it has been subject to several follow-up observations \citep{Wuyts2012StellarPopsLensed, Saintonge2013MolGasLensedGalaxies, Malhotra2017HerschelLensedCII, Gazagnes2018NeutralGasPropertiesLyC, Rigby2018MegaSauraI, Chisholm2019MegaSauraMetallicity, Solimano2021MolecularGasIntermediateMass}, becoming one of the best-studied Lyman-break galaxies (LBGs). Here, we report the discovery of an LAH associated with this galaxy and its close companion, which thanks to the lensing effect spans $\sim\ang{;;20}$ on the sky. We take advantage of its extreme  magnification (between $\mu\approx 10$ and $\mu\approx 100$ across the whole system) to spatially sample the \lya~line on subkiloparsec scales, thus offering a unique view of the CGM.
  
  Throughout this paper, we adopt a flat $\Lambda$CDM cosmology with a matter density of $\Omega_{m,\,0}=0.3$ and a Hubble parameter at $z=0$ of $H_0=\SI{70}{\kilo\meter\per\second\per\mega\parsec}$. Unless otherwise specified, we will refer to physical (proper) distances rather than comoving distances. Also, all photometric magnitudes quoted in the paper are in the AB system. When relevant, we assume a universal \citet{Chabrier2003IMF} initial mass function.

  \section{Observations and reduction}\label{sec:redux}
    \subsection{MUSE}\label{sec:muse-redux}
    We observed \arc~with the MUSE \citep{Bacon2010MUSEpaper} mounted at the Unit Telescope 4 (Yepun) of the VLT. The data were taken between 2018 April and 2019 February as part of program 101.A-0364 (PI: L\'opez), using the Wide Field Mode with adaptive optics and an extended wavelength range (WFM-AO-E). This setup provides a field of view (FoV) of $\ang{;1;} \times \ang{;1;}$ with a pixel scale of \ang{;;0.2}, and a wavelength range of \SI{4600}{\angstrom}-\SI{9350}{\angstrom} with a resolving power of $R\sim 1770$ at \SI{4800}{\angstrom} (the wavelength of \lya~emission at $z\sim2.92$). A subset of these data were already presented by \citet{Tejos2021ArctomoJ1226Kinematic}, and in this paper we follow similar reduction steps. The main difference is that \citeauthor{Tejos2021ArctomoJ1226Kinematic} only combined 20 exposures (six were discarded due to slightly suboptimal seeing) to create the final stacked data cube, while here we used the full set of 26 exposures to reach deeper into the \lya~surface brightness (SB). All exposures have an exposure time of $\sim\SI{640}{\second}$, resulting in a total exposure time of 16,640~s ($\sim4.6$~hr). We found that including these additional exposures provides a 14\% reduction in the $5\sigma$ noise level of the \lya~pseudo narrowband image, while the PSF full width at half maximum (FWHM; \SI[range-units=repeat]{0.72+-0.03}{\arcsecond}, see below) stays within the uncertainties of the value measured in the original cube (\SI[range-units=repeat]{0.76+-0.04}{\arcsecond}).
     
    The data was reduced using the MUSE pipeline \citep{Weilbacher2012MusePipeline} within the ESO Recipe Execution Tool (EsoRex) environment \citep{ESO2015CPL}, using standard procedures and calibrations. The wavelength solution was calibrated to vacuum. We measured and cleaned residual sky contamination from the standard pipeline using the Zurich Atmosphere Purge code \citep{Soto2016ZAP}. We then aligned the World Coordinate Ssytem (WCS) of the data cube to match that of the HST imaging (see below) using a bright star in the FoV as reference. The effective PSF FWHM is $\SI[range-units=repeat]{0.72+-0.03}{\arcsecond}$, as measured from a Moffat fit to the bright star in a spectrally stacked image around the wavelength of \lya~at $z=2.92$.
     
    The MUSE pipeline propagates uncertainties in the individual pixel tables during the data cube creation, yielding a ``variance cube'' as part of the output. However, variance cubes generated in this way are known to underestimate the true variances of the data, because the algorithm neglects the spatial correlation in the noise introduced by the dithering and the slicer patterns \citep[e.g.,][]{Bacon2017MuseHUDFDescription}. Since the analysis presented in this paper required a good knowledge of the uncertainties, we estimated the true variances following the method outlined by \citet{Urrutia2019MuseWideSurveyDr1} and \citet{Weilbacher2020MUSE_DPP}, but with a small modification. While they assumed the noise is spatially uniform (and hence only wavelength-dependent), we kept the spatial structure of the original variance cube, but scaled it up according to a correcting factor. The rationale behind this choice is to account for the sources' contribution to variance, which becomes relevant in the brightest regions of the LAH.  To obtain the correction factor, we created a new pixel table populated with noise following a Gaussian distribution (mean=0, variance=1), and then fed it to the \verb|muse_scipost| routine in the EsoRex pipeline to produce a mock noise data cube. We then measured the variance in sky regions of the resulting data cube to be close to 0.4 (the actual value is slightly wavelength-dependent, but we assume it is constant). So the correction factor is $1/0.4=2.5$, which is then multiplied by the original variance cube. After this correction, the signal-to-noise ratio (S/N) distribution of sky apertures is consistent with a variance of $\sim1$ while without the correction such values are $\sim2$.
    
    \begin{figure}[!htb]
        \centering
        \includegraphics[width=\columnwidth]{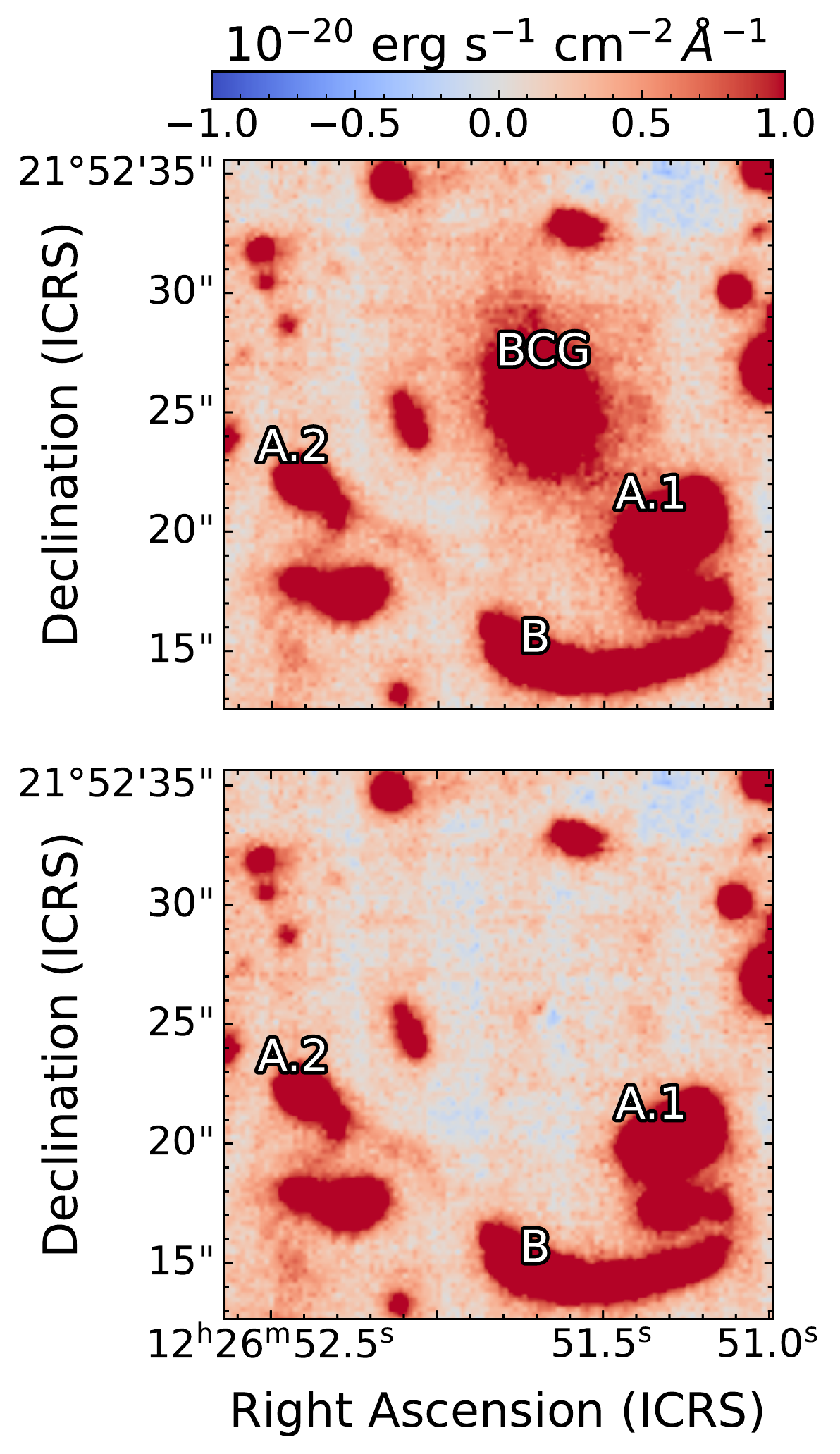}
        \caption{Subtraction of the BCG light model from the MUSE cube. The upper panel shows the inner $\ang{;;23}\times\ang{;;23}$ of an average channel map integrated from \SIrange{4900}{5700}{\angstrom} taken from the data cube. The lower panel shows the same view but the average was obtained from the BCG-subtracted residual cube created with \textsc{buddi}. The image does not contain significant features at the position of the BCG within the colormap cuts ($\pm 7\sigma$).}
        \label{fig:bcg_subtraction}
    \end{figure}
     
    After visual inspection of the data cube, it became apparent that the central galaxy of the lensing cluster (hereafter BCG) contributes significantly to the background light near the arcs (see \autoref{fig:bcg_subtraction}). Due to the position, orientation, and angular extension of the arcs with respect to the BCG, modeling and subtracting the background contamination in apertures would have been impractical and prone to many systematic effects. Instead, we opted for the self-consistent approach of modeling the BCG with a parametric light profile as a function of wavelength. We achieved this using the Bulge-Disc Decomposition of IFU Data Cubes  package \citep[\textsc{buddi}; ][]{Johnston2017BUDDI}, which uses \textsc{GalfitM} \citep{Haeussler2013MegaMorph} to model the 2D light profile of a galaxy simultaneously across several wavelength slices. While \textsc{buddi} is usually used to cleanly extract the spectra of each component included in the model, it also provides a residual data cube where the light of the target has been subtracted and the foreground and background objects can be analyzed with minimal contamination. In this case, we modeled the BCG with a single S\'ersic profile plus a central PSF component after masking the arcs and other unrelated sources. After extracting the best-fit model data cube, the residual data cube was obtained, in which the light of the arcs is free from contamination from the BCG (see bottom panel of \autoref{fig:bcg_subtraction}). In what follows, we use this BCG-subtracted data cube to perform the \lya~analysis.

    \subsection{HST}\label{sec:hst}
    Observations of \arc~were taken with the HST with the Advanced Camera for Surveys (ACS) in the broadband filters F606W and F814W, and with the infrared channel of the Wide Field Camera 3 (WFC3) in the F110W and F160W filters, as part of General Observer programs \#12368 and \#15378, respectively. 
    We used \textsc{DrizzlePac}'s  AstroDrizzle routine to align and combine the calibrated exposures to a common grid with a pixel size of \ang{;;0.03} using a gaussian kernel with \verb|pixfrac=0.8|. The images reach $5\sigma$ limiting magnitudes \footnote{With $\sigma$ computed as the standard deviation of a sample of \num{5000} flux measurements from randomly placed \ang{;;1} apertures in blank sky regions.} at $m_{606}=25.9$, $m_{814}=25.4$, $m_{110}=24.7$, and $m_{160}=24.6$. To check the accuracy of the astrometric solution \textit{a posteriori}, we crossmatched the point sources in the final ACS frames with the Gaia DR2 catalog \citep{GaiaCollaboration2018Release2} and found five sources with a mean shift of \ang{;;0.08} and no signs of rotation or major distortion.
    
\begin{figure*}[!hbt]
        \centering
        \includegraphics[width=\linewidth]{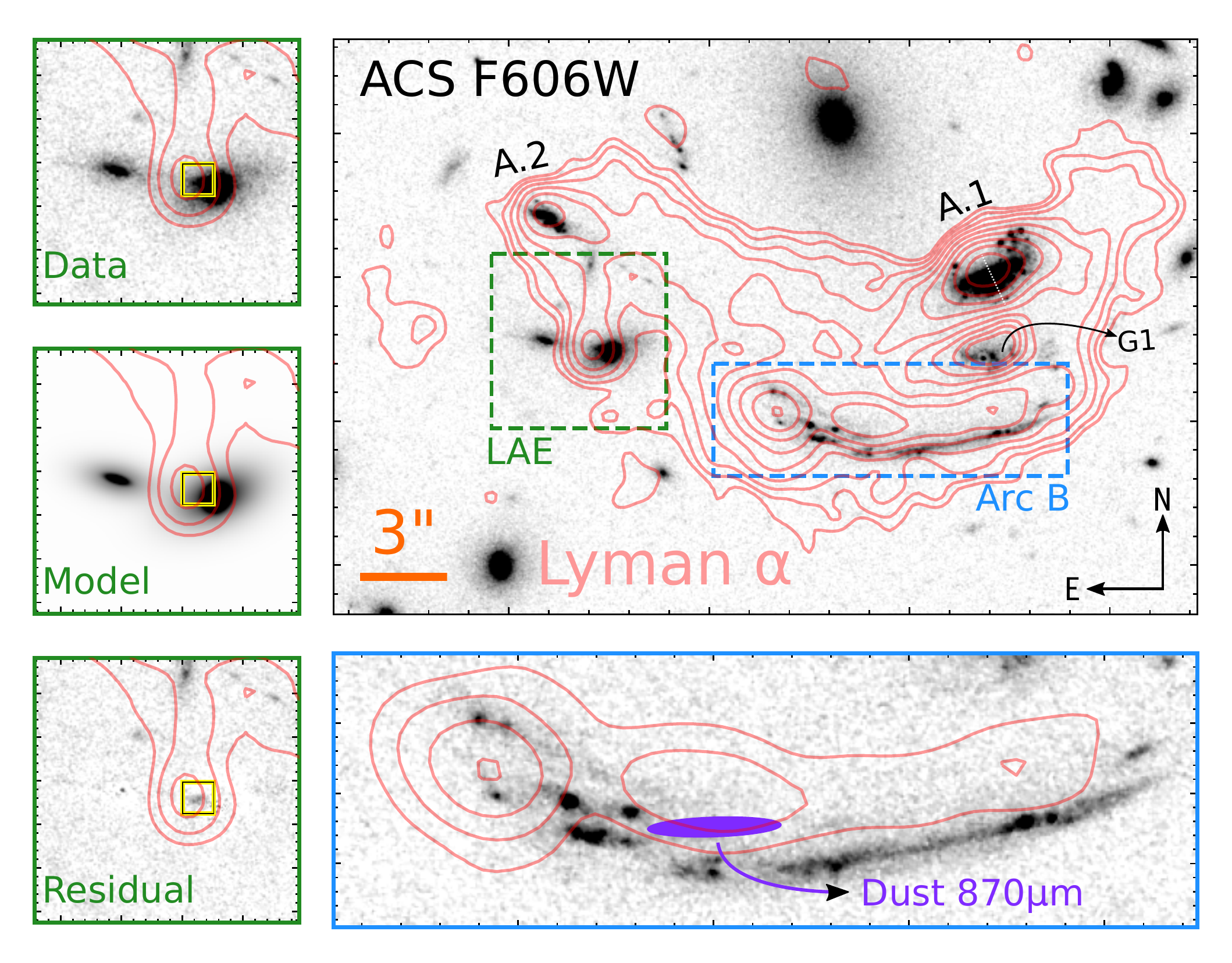}
	\caption{Overview and details of the image plane morphology of the SGASJ1226 system. Upper right: $\ang{;;30}\times\ang{;;20}$ cutout of the HST ACS F606W image, with MUSE \lya~contours overplotted (pale red). The contours start at the $3\sigma=\SI{2.27e-18}{\erg\per\second\per\centi\meter\squared\per\arcsec\squared}$  SB level of the smoothed NB image (see \autoref{sec:continuum}) and increase in powers of $\sqrt{2}$. The thin, dotted white line segment approximately divides A.1 into the two symmetric halves of the arc. Lower right: Zoomed-in view of arc B that highlights the spatial offset between the UV and \lya~emission. For clarity, contours only show the top $12\sqrt{2}, 24, 24\sqrt{2}$ and $48\sigma$ levels of \lya~SB. Top left: \ang{;;6.1} cutout of the ACS image centered on a local maximum of \lya~SB, with contours starting at $6\sigma$. Middle left: \textsc{Galfit} model of the two foreground elliptical galaxies near the local \lya~peak. Lower left: Residuals from the subtraction of the \textsc{Galfit} model. A compact excess of continuum appears near the center of the \lya~peak. The purple ellipse shows the location and deconvolved size of the dust continuum emission at \SI{870}{\micro\meter} detected with the Atacama Compact Array \citep[ACA;][]{Solimano2021MolecularGasIntermediateMass}}
        \label{fig:hst_lya_contours}
    \end{figure*}

    A $\ang{;;30}\times\ang{;;20}$ cutout of the ACS F606W image is shown in the upper right panel of \autoref{fig:hst_lya_contours}. The multiple images and arcs are labeled using the same nomenclature in \autoref{fig:bcg_subtraction}. A.1 is the brightest (most highly magnified) component and corresponds to a twofold, almost symmetric pair of images of galaxy A on both sides of the lensing critical curve (dotted white line). A fainter counterimage (A.2) of the same galaxy is seen $\sim\ang{;;15}$ to the east. The second galaxy of the system, B, has a single known image (arc B) stretching from east to west a few arcseconds below A.1. A zoomed-in view of arc B shown in the lower right panel of \autoref{fig:hst_lya_contours} reveals multiple knots of UV emission. Arc B is also the host of the unresolved \SI{870}{\micro\meter} continuum detection (purple ellipse) reported in \citet{Solimano2021MolecularGasIntermediateMass}. Between arcs A.1 and B we indicate G1, the $z=0.77$ foreground \ion{Mg}{2} absorber studied in \citet{Mortensen2021ScoopArctomoJ1226} and \citet{Tejos2021ArctomoJ1226Kinematic}.
    
    \begin{table}[!htb]
        \centering
        \caption{HST Aperture Photometry of the $z=2.92$ Lensed Aucs of the \name~System.}
        \begin{tabular}{cccc}
            \hline
                    & \multicolumn{3}{c}{AB Magnitude\tablenotemark{a}} \\
            Filter/Image & A.1\tablenotemark{b} & A.2 & B.1   \\
             \hline
             ACS F606W & \num{20.62+-0.04} & \num{22.29+-0.05} & \num{21.27+-0.12} \\
             ACS F814W & \num{20.47+-0.04} & \num{22.15+-0.06} & \num{21.00+-0.11} \\
             WFC3 F110W & \num{20.51+-0.04} & \num{22.23+-0.08} & \num{20.94 +-0.12}\\
             WFC3 F160W & \num{20.13+-0.04}  & \num{21.91+-0.09} & \num{20.46 +-0.12} \\
             \hline
             \hline
        \end{tabular}
        \raggedright
        \tablenotetext{a}{Corrected by Galactic extinction but not by lensing magnification.}
        \tablenotetext{b}{Includes both halves of the arc. Contamination from the lens perturber is not accounted for.}
        \label{tab:hst_photometry}
    \end{table}
    
    At $z=2.92$ the observed filters cover the UV continuum and the Balmer break at the source rest frame, sampling  the spectral energy distribution (SED) of the young stellar populations. The high resolution imaging also provides exquisite morphological detail on the lensed arcs, revealing numerous clumps and substructures. At the same time, it enables the identification of lensed image pairs  and the positions of cluster members that are included in the development of the  lens model \citep{Dai2020AsymmetricBrightnessJ1226Arc, Tejos2021ArctomoJ1226Kinematic}.
    
    The complex morphology of the lensed galaxies in the \name~system (see \autoref{sec:radial_profiles}) makes the standard photometric extraction techniques unsuitable. For this reason, we used manually defined ad hoc polygonal apertures to obtain the total flux from the arcs. To assess the systematic error associated with this method, four members of the team created several apertures for each arc based on the image with the broadest PSF (WFC3-IR/F160W). The flux standard deviation derived from 20 different contributed apertures is about 10\%, three times larger than the statistical error inferred from linear propagation. The final image plane magnitudes are reported in \autoref{tab:hst_photometry} after correction from Milky Way reddening using the \citet{SchlaflyFinkbeinerMWReddening} extinction tables.

    \subsection{Lens model}\label{sec:lens}
     A proper measurement of the intrinsic properties of strongly lensed galaxies requires accurate knowledge of the geometrical distortion and magnification produced by the lensing cluster. Good approximations of the amount and direction of the distortion at any given position are obtained by modeling the cluster as a collection of parametric dark matter profiles and then constraining the model to predict the positions of image pairs identified in the data.  In this paper, we use the lens model presented in \citet{Tejos2021ArctomoJ1226Kinematic}, and therefore we refer the reader to that work for further details. Briefly, the modeling was done using the \verb|Lenstool| software \citep{Jullo&Kneib2009LensTool} following the  procedure described in \citet{Sharon2020ClusterLensModelsSGAS}. The model was fitted using three images of galaxy A (two images in A.1 plus the counterimage A.2; see \autoref{fig:hst_lya_contours}), with the position of individual clumps identified in the HST data serving as constraints. Both cluster-scale and galaxy-scale potentials were modeled as pseudoisothermal ellipsoidal mass distributions (PIEMDs), all located at $z_\text{lens} = 0.43$. For the galaxy-scale potentials the center of each PIEMD was held fixed to match the optical centroids of the cluster members during the fitting process. The presence of a second strong-lensing cluster at the same redshift only \ang{;157;} to the south of \name~motivated the addition of a cluster-scale PIEMD at that position, contributing shear to the overall lensing potential. Finally, a perturber on top of the western part of arc A.1 and the $z=0.77$ galaxy between A.1 and B \citep[labeled G1 in][ see \autoref{fig:hst_lya_contours}]{Tejos2021ArctomoJ1226Kinematic} were included as individual components. The latter was treated as belonging to the same plane of the foreground cluster.

     The resulting best-fit model provides a set of deflection matrices that prescribe the angular offsets (in both the R.A. and decl. axes) induced by the lens at any given position in the image plane. This model reproduces the positions of 26 constraints with an rms of \ang{;;0.08}. Throughout this paper, we use the deflection matrices to reconstruct source plane positions and sizes.

     In \autoref{fig:smooth_nb} we plot the lensing critical curve (i.e., the locus of maximal magnification) on top of the MUSE pseudo-NB image as a way to visualize the morphology of the lensing potential.

  \section{Results}\label{sec:results}
    \subsection{Host galaxy properties}\label{sec:sed}
    In this section we give a characterization of the host galaxies in terms oftheir systemic velocity, mass, and luminosity.
    A crucial step to understanding the origins and kinematics of the \lya-emitting gas is to secure the systemic redshifts for the galaxies within the halo. For galaxy A, numerous reports exist in the literature. First, in their discovery paper \citet{Koester2010ArcJ1226Discovery} used UV absorption lines to get $z_\text{ISM}=2.9233$. Later, \citet{Wuyts2012StellarPopsLensed} obtained Keck/NIRSPEC spectroscopy that allowed the measurement of rest-frame optical nebular lines to find $z_\text{neb}=\num{2.9257+-0.0004}$. The difference of $\sim\SI{200}{\kilo\meter\per\second}$ between the emission and absorption solutions was then suggested as tentative evidence for outflows \citep{Wuyts2012StellarPopsLensed}. For galaxy B, Gemini/GMOS spectroscopy has yielded $z_\text{ISM}=2.9233$ \citep{Bayliss2011GMOSLensing} but no nebular redshift is available. Here, we avoid the systematic effects arising from a mix of different instruments by measuring redshifts directly on the MUSE data cube. Unfortunately, the nebular lines accessible in the MUSE wavelength range are extremely faint, so we coadded all spaxels associated with each galaxy to improve the S/N, thereby losing any information on possible spatial velocity gradients. For each galaxy, we simultaneously fit the \ion{Si}{2}* $\lambda 1533$, [\ion{C}{3}] $\lambda 1906$ and \ion{C}{3}] $\lambda 1908$ lines on the continuum-subtracted spectrum with Gaussian profiles of a common width, varying only $z$ and the line amplitudes. In this way we obtained $z_\text{neb}=\num{2.9257+-0.0001}$ and $z_\text{neb} = \num{2.9238+-0.0002}$ for A and B, respectively, corresponding to a radial velocity difference between A and B of \SI{145+-17}{\kilo\meter\per\second}. The value for galaxy A is fully consistent with prior literature measurements \citep{Wuyts2012StellarPopsLensed, Rigby2018MegaSauraI}. We checked that our solutions are robust against the relaxation of the width constraint: letting $\text{\ion{Si}{2}}^{*}\,\lambda 1533$  have a width parameter independent of the \ion{C}{3} lines yields the same 16\textsuperscript{th}-50\textsuperscript{th}-84\textsuperscript{th} percentiles as the single-width run.
    
    \begin{table}[!htb]
        \centering
        \caption{Properties of the two largest member galaxies}\label{tab:properties}
        \begin{tabular}{lcc}
            \tableline
             Property & Galaxy A & Galaxy B \\
             \tableline
             Redshift\tablenotemark{a} & \num[separate-uncertainty=false]{2.9257+-0.0001} & \num[separate-uncertainty=false]{2.9238+-0.0002} \\
             Average $\mu$ & \num{7.5+-1.5}\tablenotemark{b} & \num{30+-6} \\
             $\log\left(\text{M}_\text{stars}/\text{M}_\odot\right)$ & \num{9.8+-0.2}\tablenotemark{b} & \num{9.7+-0.2} \\
             SFR $(\text{M}_\odot\,\text{yr}^{-1})$\tablenotemark{c} &  \num{10+-2}\tablenotemark{b} & \num{15+-4} \\
             $\log\left(L_\text{UV}/L_\odot\right)$ & \num{10.80+-0.09}\tablenotemark{b} & \num{10.6+-0.1} \\
             $\log\left(L_\text{IR}/L_\odot\right)$\tablenotemark{c} & $< \num{11.5}$ & \num{10.9+-0.3}\\
             $\beta_\text{UV}$ & \num{-0.82+-0.15} & \num{-0.66+-0.24} \\
             $12+\log\left(\text{O}/\text{H}\right)$\tablenotemark{d} & \num{8.2+-0.2} & \num{8.2+-0.2} \\
             \tableline
        \end{tabular}
        \raggedright
        \tablenotetext{a}{Based on the simultaneous fit to the $\text{\ion{Si}{2}}^{*}\,\lambda$1533  and $\text{\ion{C}{3}]}\,\lambda\lambda$1906, 1908 nebular emission lines in the MUSE spectra.}
        \tablenotetext{b}{Estimated from the complete lensed counterimage A.2; see \autoref{sec:lens} and \autoref{sec:sed}. The average magnification of the arc A.1 is $\mu \approx 87$.}
        \tablenotetext{c}{From \citet{Solimano2021MolecularGasIntermediateMass}.}
        \tablenotetext{d}{From \citet{Chisholm2019MegaSauraMetallicity} \textsc{Starburst99} fits assuming stellar metallicity equals gas-phase metallicity.}
    \end{table}
    
    The general properties of the two brightest galaxies in the LAH were already reported in \citet{Solimano2021MolecularGasIntermediateMass} by means of fitting SED models to the available broadband photometry. Briefly, the SED fits included photometry from the four HST bands mentioned above, Spitzer/IRAC 3.6 and \SI{4.5}{\micro\meter} bands, and the ACA \SI{870}{\micro\meter} continuum. The available Herschel bands were excluded due to severe blending and low S/N \citep[see][]{Saintonge2013MolGasLensedGalaxies}. The magnified fluxes were then fit with the \textsc{Magphys} \citep{daCunha2008MagphysPaper, daCunha2015HighZMagphysALESS} energy balance code to constrain the star formation rate (SFR), stellar mass, and infrared luminosity. The lack of midinfrared bands prevented the detection of continuum from dust heated by an active galactic nucleus (AGN). Here, we reproduce the best-fit values and their uncertainties after demagnification in \autoref{tab:properties}. The average magnification was computed as the ratio between the solid angle of the HST aperture in the image plane and the solid angle spanned by the smallest polygon that encloses all the delensed grid points in the source plane. The magnification uncertainty was fixed to 20\% to account for typical systematic errors in the lens modeling \citep[][where statistical errors are less important]{Raney2020FrontierFieldsLensingUncertainties}. We tested this choice by computing the actual statistical uncertainties using a set of 100 realizations of the lens model sampled from a Markov Chain Monte Carlo (MCMC) chain. We found that for the apertures used here, the relative uncertainty in magnification is between 13\% and 17\%.

    We also estimated the rest-frame far-UV luminosity of the arcs directly from the ACS/F606W photometry tracing the stellar continuum at $\lambda_\text{rest}\approx\SI{1500}{\angstrom}$ at $z=2.92$. As we did with all the available HST data, we divided the image plane fluxes by the average magnification factor of the most complete image. This means that despite A.1 having the largest S/N, our flux-dependent estimates for galaxy A come from the counterimage A.2 instead, since it has the largest footprint in the source plane. In this way, we used $\mu=7.5$ for A.2 and $\mu=30.2$ for the arc B to find $\log(L_\text{UV}/L_\odot) = \log(\lambda L_\lambda) = \num{10.80+-0.09}$ for galaxy A and $\log(L_\text{UV}/L_\odot) = \num{10.6+-0.1}$ for galaxy B, corresponding, respectively, to \num{1.0+-0.2}  and \num{0.64 +- 0.15} times the typical UV luminosity at $z=3$,  $L^{*}=\SI{6.2e10}{\lumsol}$  \citep[e.g.,][]{Paltani2007LuminosityFunctionUV, Reddy2008LuminosityFunctionUV, Bian2013LuminosityFunctionUV, Mehta2017LuminosityFunctionUV}. 
    
    \subsection{Continuum Subtraction and Pseudo-NB Imaging}\label{sec:continuum}
    The first step in our analysis was to create NB \lya~images from the reduced MUSE cube. We first extracted a subcube between \num{4754} and \SI{4810.3}{\angstrom}, a range that fully includes both the \lya~emission line and \SI{10}{\angstrom} of the adjacent continuum on each side. In the UV-bright regions of the arc, the continuum shows a clear break at the wavelength of \lya. At bluer wavelengths the continuum is strongly suppressed by a combination of damped self-absorption and decreased IGM transmission, while at redder wavelengths the emission is dominated by the UV power-law continuum but modulated by the \ion{H}{1} damping wings. These effects produce an underlying continuum with a complex shape. In principle, one could model it as the superposition of the UV power law, a damped Voigt profile, and the break from the IGM transmission; however, the data lacks sufficient S/N to fit such a model on a spaxel-to-spaxel basis. Instead, we chose to model the continuum as a linear ramp between the blue and red continuum levels. This scheme has been successfully applied to similar datasets \citep[e.g.,][]{Claeyssens2019LyaSpectralVariations}. The functional form of this model is
    
    \begin{equation*}
        f_\lambda = \begin{cases}
        f_\text{blue} & \SI{4754}{\angstrom} \leq \lambda < \lambda_1 \,,\\
        m(\lambda - \lambda_1) + f_\text{blue} & \lambda_1 \leq \lambda \leq \lambda_2\,, \\
        f_\text{red} & \lambda > \lambda_2 \geq \SI{4810.3}{\angstrom}\,,
        \end{cases}
    \end{equation*}
    
    \noindent where $m=(f_\text{red} - f_\text{blue})/(\lambda_2 - \lambda_1)$ is the slope of the linear ramp. We estimated the mean blue ($f_\text{blue}$) and red ($f_\text{red}$) continuum levels by averaging the spectral channels to each side of the line from the subcube, at $\lambda < \lambda_1$ and $\lambda > \lambda_2$ respectively. We found that the values $\lambda_1=\SI{4768.8}{\angstrom}$ and $\lambda_2=\SI{4788.0}{\angstrom}$ produce robust continuum subtraction (average zero flux) on both sides of the line.

\begin{figure*}
        \centering
        \includegraphics[width=\linewidth]{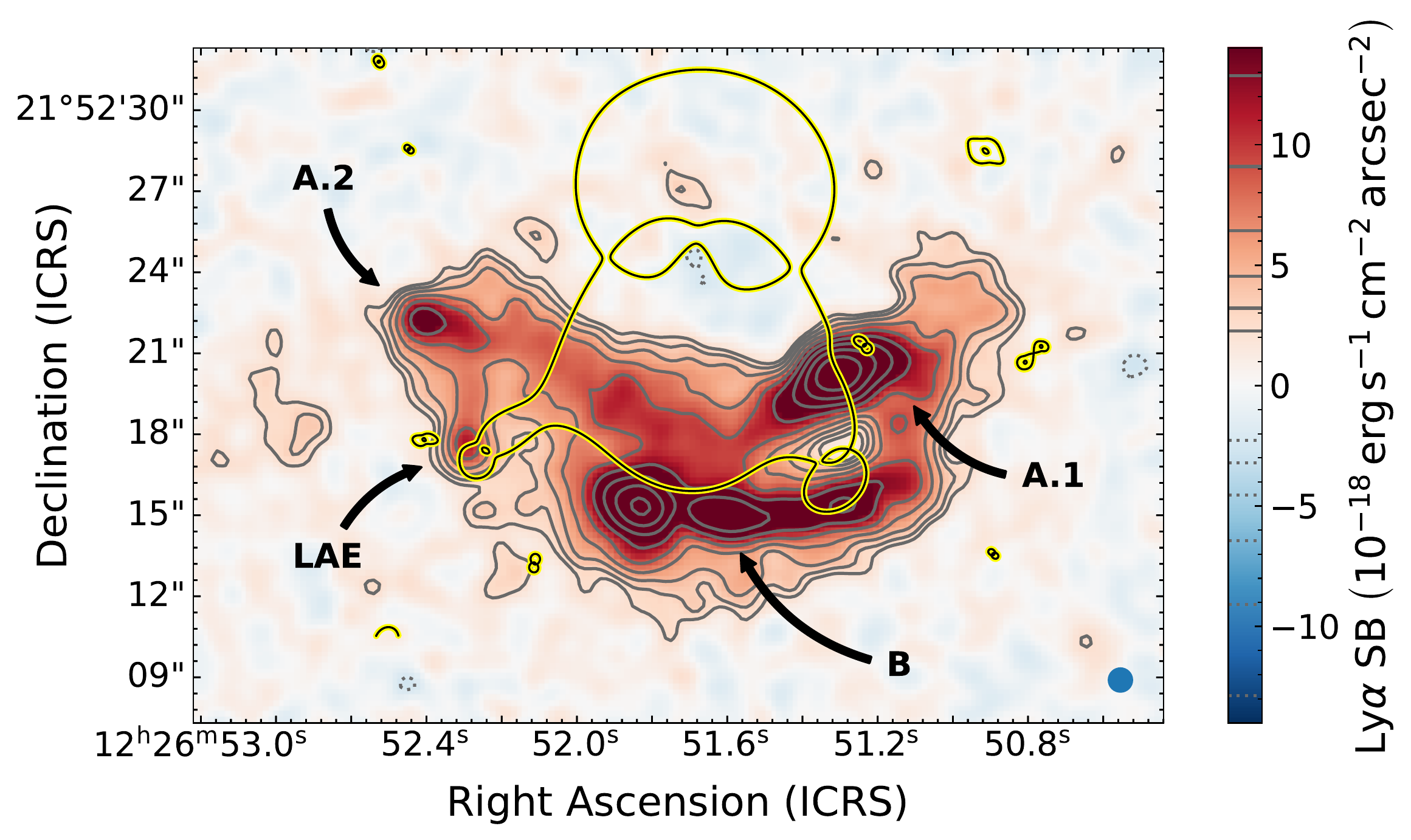}
        \caption{Continuum-subtracted pseudo-NB image of the lensed LAH at $z=2.92$. The image was smoothed with a gaussian kernel of 2 spaxels' width. The blue circle at the lower right corner indicates the FWHM size of the smoothing kernel, setting the angular resolution at effectively \ang{;;0.94}.  The positive (negative) contours shown as gray solid (dotted) lines are the same as those in \autoref{fig:hst_lya_contours} and start at the $\pm 3\sigma=\pm\SI{2.27e-18}{\erg\per\second\per\centi\meter\squared\per\arcsec\squared}$ level of SB  and increase (decrease) in powers of $\sqrt{2}$. The black curve with a yellow outline traces the lensing critical curve, that is, the locus of maximum magnification.}
        \label{fig:smooth_nb}
    \end{figure*}

    Once the continuum was extracted from every spaxel, we integrated the resulting cube between \num{4769} and \SI{4788}{\angstrom} (rest frame \num{1215.3} and \SI{1219.6}{\angstrom} at $z=2.924$) to obtain a pseudo-NB image of the \lya~emission. The integration limits were chosen to enclose the full spectral extent of the redshifted line. We excluded the blueshifted peak since it only appears in a limited image plane region and its contribution to the total flux is less than 1\%. \autoref{fig:smooth_nb} shows the final image after smoothing with a Gaussian kernel of $\sigma=2$ spaxels. The smoothing was applied only for the ease of visualizing the low-SB structure of the object, but in the subsequent analysis we used the unsmoothed version.
    
    \subsection{Image Plane Analysis}\label{sec:img-plane}
    In order to obtain insights into the spatial \lya~properties of \arc~in a way that is independent of the lens model, we started our analysis in the image plane, rather than in the source plane. The NB image in \autoref{fig:smooth_nb} shows at least five strong peaks of \lya~SB that stand out from the diffuse emission. In aid of comparing these features to the UV continuum, we reproduce the \lya~SB contours of \autoref{fig:smooth_nb}  in \autoref{fig:hst_lya_contours}, placing them on top of the ACS F606W image, the band that traces the UV continuum at $z=2.92$. The two northernmost local peaks are associated with images A.1 (west, bright) and A.2 (east, faint) of galaxy A. Toward the south, we observe the second brightest peak of \lya~SB, which is connected to a fainter arc-like structure extending about \ang{;;5} to the east. The arc itself has two secondary peaks at the $\approx 34\sigma$ level. We associate the peak plus the arc with Arc B in the HST image, although there is an evident offset between the bright spots in \lya~and the location of the UV-bright clumps (see lower right panel of \autoref{fig:hst_lya_contours}). In particular, the bright \lya~peak at B is offset by \ang{;;1.2} with respect to the UV centroid, which lies in the middle of the brightest knots of the arc. Also, the \lya~arc is offset by \ang{;;0.6} on average to the north of the UV arc B. Due to the achromatic nature of lensing, offsets in the image plane between two tracers (e.g., \lya~and UV here) imply intrinsic offsets in the source plane.
    
    The other remarkable local peak is $\sim \ang{;;5}$ southward from A.2 at $17\sigma $ above the sky background (see green box in \autoref{fig:hst_lya_contours}). Interestingly, this peak is centered very close to two intervening foreground cluster members. Since no emission line is expected at $\lambda_\text{obs}=\SI{4471}{\angstrom}$ at the redshift of the cluster ($z_\text{lens} = 0.43$), we conclude that the signal comes from the same redshift of the LAH and is likely originates in a compact \lya~emitter (LAE) embedded in the halo (hereafter we refer to this \lya~source as \arc-LAE). Motivated by this hypothesis, we searched the HST data for a continuum counterpart, finding a candidate near the core of one of the cluster members in the F606W image. To confirm the presence of this counterpart, we employed \textsc{Galfit} \citep{Peng2002Galfit1, Peng2010Galfit3} to model the light of the two intervening galaxies with S\'ersic profiles. After subtraction of the best-fit model, the residuals clearly show an excess emission within \ang{;;0.2} of the \lya~peak. The excess is also detected in the F814W image, but not in the near infrared filters. The limited spatial resolution in the F110W and F160W bands results in contamination the expected location of the galaxy by PSF subtraction residuals, preventing us from putting any constraint on its flux.

    We measured the flux in the residual images with SExtractor's \citep{Bertin1996SExtractor} automatic apertures with the F606W residual as the detection frame, yielding (magnified) magnitudes of $m_{606}=\num{25.94+-0.17}$ and $m_{814}=\num{25.82+-0.30}$. From these two bands we inferred a UV slope of $\beta=\num{-1.6+-1.1}$. A \lya~flux of \arc-LAE of $\SI{7.5e-17}{\erg\per\second\per\cm\squared}\mu^{-1}$  was measured by fitting an asymmetric Gaussian (AG) profile (see \autoref{sec:asym-gauss}) to the spectrum integrated in a \ang{;;1} circular aperture. Extrapolating the continuum flux to $\lambda_\text{rest}=\SI{1215.67}{\angstrom}$ we estimated a rest-frame \lya~equivalent width (EW) of \SI{104+-19}{\angstrom}, which falls in the classical definition of high-redshift LAEs \citep{Ouchi2020LymanAlphaReview}. At the same redshift as the rest of \name, this source is likely a satellite galaxy of the system. Details on the characterization of this source together with estimates of its contribution to the total \lya~luminosity  are presented in \autoref{sec:satellite_lae}. 
    
    \subsubsection{Integrated spectrum}\label{sec:integrated_spectrum}
\begin{figure}[!hbt]
        \centering
        \includegraphics[width=\columnwidth]{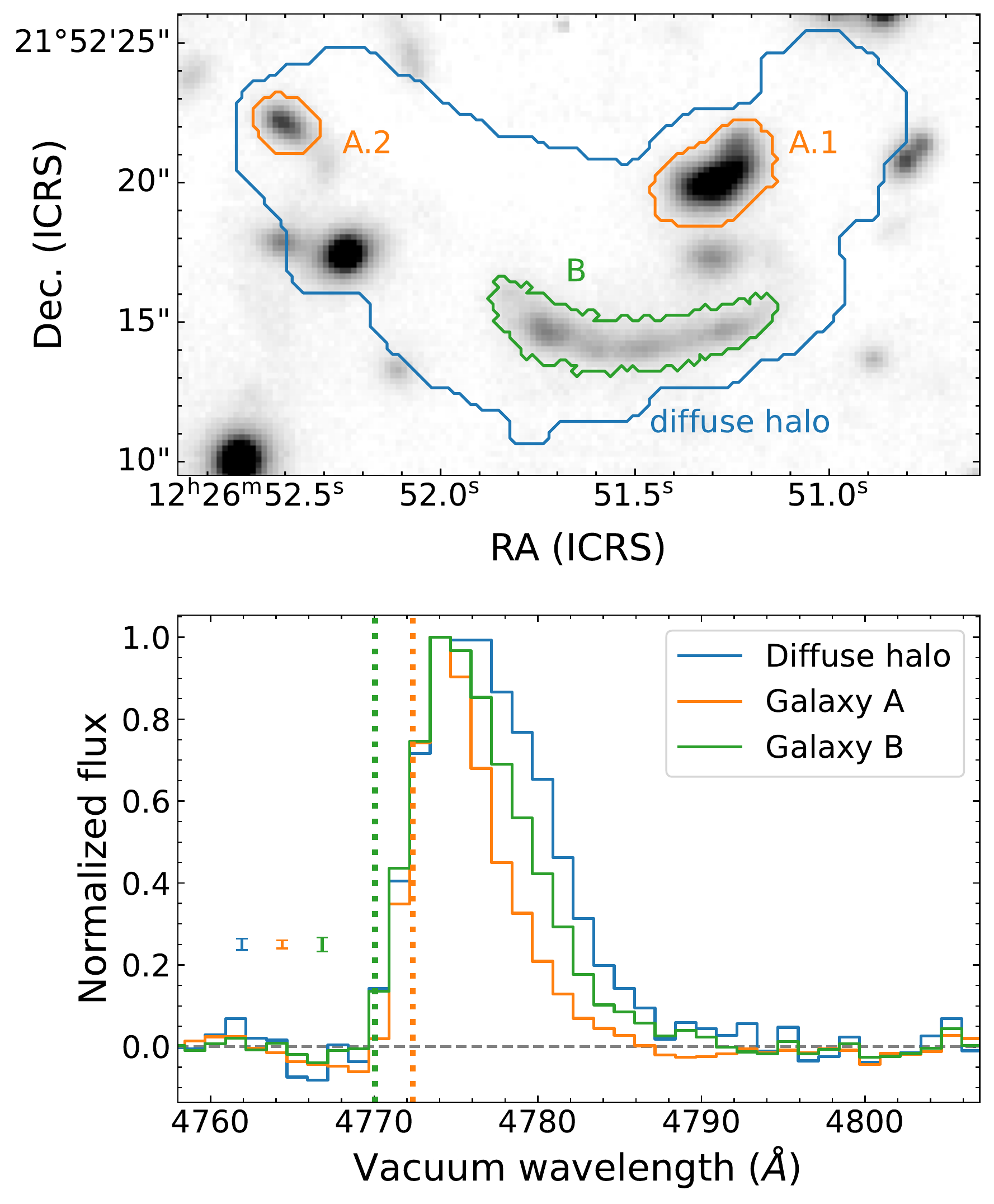}
        \caption{Top: MUSE broadband image synthesized around rest-frame \SI{1600}{\angstrom} with the global image plane apertures used for extraction shown as colored lines. Bottom: Continuum-subtracted, normalized MUSE spectra from three different image plane apertures. The blue, orange, and green curves correspond to the spectra of the diffuse halo, galaxy A, and galaxy B, respectively. Representative error bars are shown on the left. The dotted orange and green vertical lines indicate the wavelength of \lya~at the systemic redshifts (see \autoref{sec:sed}) of galaxy A and galaxy B, respectively.}
        \label{fig:global_profiles}
    \end{figure}

    In \autoref{fig:global_profiles} we show the spatially integrated \lya~spectrum for three different apertures, one for each of the two arcs and another for the diffuse emission. To create the apertures, we first integrated the data cube at $\SI{1550}{\angstrom} < \lambda_\text{rest} <  \SI{1650}{\angstrom}$ to obtain a map of the UV continuum at $z=2.92$ at the same resolution, pixel scale, and astrometry of the \lya~NB image. Then, we applied a threshold of continuum $\text{S}/\text{N}=5$ on this image to isolate spaxels containing bright UV emission. The resulting masks were manually inspected and tweaked to remove spurious spaxels that met the S/N threshold but were associated with unrelated sources. For galaxy A, the masks include all spaxels from both the arc A.1 and the counterimage A.2. The mask for the diffuse LAH was defined as all spaxels with \lya~SB above $3\sigma(\text{smooth})=\SI{2.2e-18}{\erg\per\second\per\centi\meter\squared\per\arcsec\squared}$ in the smoothed narrowband image. Then, the continuum masks were subtracted from the diffuse halo mask to exclude spaxels with continuum emission. The three apertures and their corresponding \lya~spectra are shown in \autoref{fig:global_profiles}.
    
    The three profiles show a very strong red peak with no clear blue component. Also, the line profile is markedly asymmetric in the three cases, with a sharp drop from the peak to the blue and a broad red wing. The major difference between the three spectra is the width of the line and the location of the peak. We measured line properties such as the FWHM and the shift of the peak with respect to the systemic velocity of the system by fitting an AG profile to the spectrum after taking into account the line spread function (LSF; for details on this method see \autoref{sec:asym-gauss}). Galaxy A shows the narrowest line, with an observed-frame FWHM of \SI[multi-part-units=single]{4.68+-0.6}{\angstrom} (equivalent to a rest-frame velocity of \SI[multi-part-units=single]{294.2+-4}{\kilo\meter\per\second}) and a \SI[multi-part-units=single]{1.72+-0.03}{\angstrom} (\SI[multi-part-units=single]{108.2+-1.8}{\kilo\meter\per\second}) offset between the peak and the systemic redshift $z_\text{sys}^A=2.9257$, indicated by a dotted orange line. The \lya~profile emitted by galaxy B is broader (FWHM = \SI[multi-part-units=single]{6.84+-0.1}{\angstrom},  \SI[multi-part-units=single]{430.1+-6.5}{\kilo\meter\per\second}), and, despite its having the same peak wavelength as galaxy A, the systemic redshift is lower ($z_\text{sys}^B=2.9238$; dotted green line) and thus the velocity offset is \SI[multi-part-units=single]{4.63+-0.06}{\angstrom} (\SI[multi-part-units=single]{291.5+-3.5}{\kilo\meter\per\second}). Finally, the diffuse halo emission has the broadest line width (\SI[multi-part-units=single]{8.10+-0.07}{\angstrom}, \SI[multi-part-units=single]{509+-4}{\kilo\meter\per\second}) and a peak that is slightly displaced redward with respect to the peaks of A and B (\SI[multi-part-units=single]{4.31+-0.05}{\angstrom}, or \SI[multi-part-units=single]{270.6+-3.0}{\kilo\meter\per\second} if we assume $(z_\text{sys}^A + z_\text{sys}^B)/2 = 2.92475$). We warn the reader that the relatively low uncertainties on the fitted parameters reflect the high S/N of the spectra, and thus do not include the systematic uncertainties arising from the unknown intrinsic profile shape.
    
    The broader and redder line profile we observe in the diffuse halo relative to the central component agrees with previous findings by \citet{Claeyssens2019LyaSpectralVariations} for two other lensed LAHs (SMACS2031 and MACS0940) and by \citet{Leclercq2020SpatiallyResolvedLyaHalosMHUDF} in a sample of unlensed LAHs in the UDF. In a scenario where the extended \lya~emission is explained exclusively by neutral gas scattering, the width of the line is linked to the number of scatterings those photons had to experience before escaping the halo in the direction of the observer. The broader the line, the more reprocessed are the photons in the high-velocity tail. In this context, our results would indicate that \lya~photons coming ``down the barrel'' from the SFGs are less reprocessed on average than photons that escape from the outskirts of the halo. Such an  effect can naturally arise if the close environments of the galaxies have a higher ionization fraction that effectively reduces the optical depth along the line of sight. However, the \lya~signal depends not only on the neutral hydrogen column density, but also on the gas kinematics.

    \subsection{Source Plane Analysis}\label{sec:src-plane}
    \subsubsection{Morphology} \label{sec:radial_profiles}
    In this section we present the spatial properties of the \lya~SB in the source plane. As we discussed in \autoref{sec:img-plane}, the NB data in the image plane already reveals that the \lya~emission is both spatially offset and more extended than the UV continuum. Qualitatively, these properties should be preserved by the lensing, but we now verify it in a quantitative way using our lens model. In what follows, we use the deflection matrices resampled to MUSE resolution (\ang{;;0.2}) unless otherwise specified.
    
    We employed a Bayesian forward-modeling approach similar to the one presented in \citet{Claeyssens2022LlamasPaper1}. We modeled both the MUSE UV and \lya~data for each of the two galaxies (A and B) as S\'ersic profiles and their parameter space was explored by a MCMC sampling scheme using the \verb|emcee| library \citep{ForemanMackey2013emcee}. For each proposed set of parameters in the chain, the model was evaluated and traced back to the image plane according to the lens model prescription, convolved with the PSF, and compared to the data under a Gaussian likelihood. For the sake of simplicity, we used a single S\'ersic profile per component with all its six parameters free and set uniform priors.
    
    We first fit the MUSE UV continuum image at $\lambda_\text{rest}\sim\SI{1600}{\angstrom}$ defined in \autoref{sec:integrated_spectrum}. We chose to fit the UV model in the MUSE data rather than the ACS F606W data to have a more comparable data quality between the UV and \lya~datasets. After MCMC convergence we obtained a best fit S\'ersic index and circularized effective radius of $n_\text{UV}^A = \num{0.57+-0.04}$ and $r_{50,\,\text{UV}}^A=\SI{0.97+-0.02}{\kilo\parsec}$, respectively, for galaxy A, while for galaxy B, $n_\text{UV}^B=\num{1.4+-0.2}$ and $r_{50,\,\text{UV}}=\SI{4.6+-0.5}{\kilo\parsec}$. For completeness and direct comparison with \citet{Claeyssens2022LlamasPaper1}, we also quote the 90\%-light radius $r_{90,\,\text{UV}}^A=\SI{1.85+-0.05}{\kilo\parsec}$ and $r_{90,\,UV}^B = 12.4_{-1.9}^{+2.2}\,\si{\kilo\parsec}$. The resulting median distance\footnote{Assumes an uncertainty of 10\% associated with the square root of magnification.} between the centers of A and B is \SI{14.3+-1.4}{\kilo\parsec}.
    
    We repeated this exercise with the \lya~NB data, finding $n_{\lya}^{A}=5.18_{-0.06}^{+0.02}$ and $r_{50,\,\lya}^A=9.3_{-0.1}^{+0.3}\,\si{\kilo\parsec}$ for galaxy A and $n_{\lya}^{B}=3.73_{-0.05}^{+0.06}$ and $r_{50,\,\lya}^B=19.4_{-0.2}^{+0.3}\,\si{\kilo\parsec}$ for galaxy B. The corresponding 90\%-light radii are $r_{90,\,\lya}^A=65.4_{-1}^{+2}\,\si{\kilo\parsec}$ and $r_{90,\,\lya}^B=\SI{101+-1}{\kilo\parsec}$. 
    
\begin{figure*}[!hbt]
        \centering
        \includegraphics[width=\linewidth]{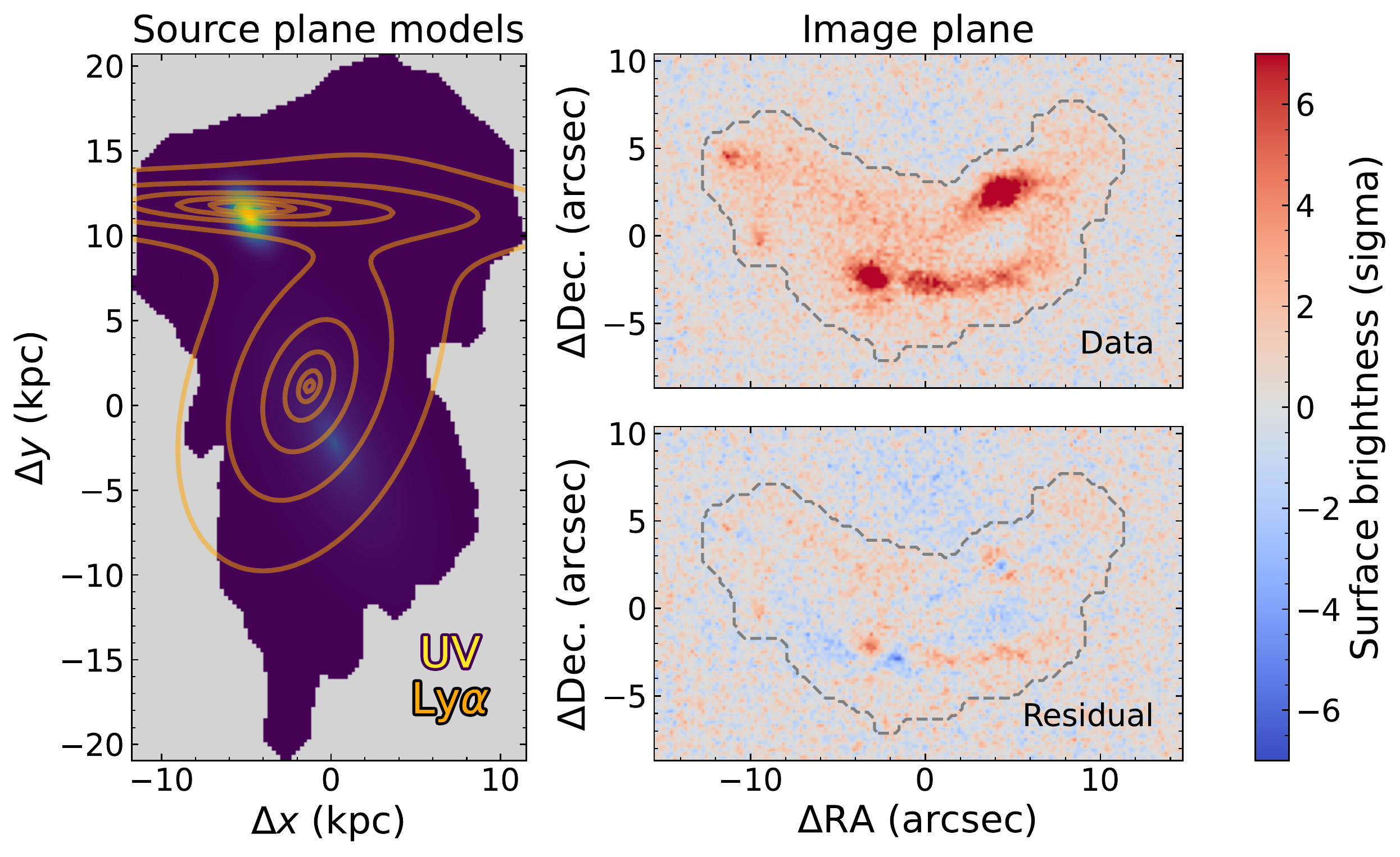}
        \caption{Left: Source plane MCMC-averaged light distribution for the UV (background image) and the \lya~emission (orange contours). The image-masked borders delimit the delensed source plane area of the isophotal mask shown in dashed lines in the right panels. The \lya~model contours end at the maximum and decrease in powers of 2. Upper left: Unsmoothed \lya~NB image in units of multiples of the background RMS. The dashed line marks the isophotal mask corresponding to the ``diffuse halo'' defined in \autoref{sec:integrated_spectrum}. Lower right: Residual from subtracting the best PSF-convolved \lya~model with the observed \lya~NB in units of the background rms (same as above).}\label{fig:sersic_src}
    \end{figure*}

    In both the UV and \lya~cases the models cannot fully reproduce all of the image plane features, as revealed by the high-significance residuals of the fit (see \autoref{fig:residual_uv} for the UV and \autoref{fig:sersic_src} for \lya). This can be a result of the clumpy nature of the galaxies and their halos, making the S\'ersic profile unsuitable. A full morphological analysis of the individual clumps is outside the scope of this paper and will be presented elsewhere. Nevertheless, the fitted S\'ersic parameters can be informative of the sizes and overall light distribution of the galaxies. For example, the high S\'ersic indices indicate that the sources have a very compact core and extended tails, similar to the double-exponential profiles often invoked for describing LAHs \citep[e.g.,]{Wisotzki2016ExtendedLyaHalosMUSE, Leclerq2017MuseHudfLyaHaloes}. The \lya~half-light radii, on the other hand, put the two sources in the top 10\% of the \citet{Leclerq2017MuseHudfLyaHaloes} sample and above any measurement in the \citet{Claeyssens2022LlamasPaper1} sample. In this context, the size of the \name~system approaches the lower end of the size range of \lya~blobs \citep{Ouchi2020LymanAlphaReview}.
    
    Using the centroids of the models, we computed the intrinsic spatial offsets between \lya~and the UV to be $\Delta(\lya-\text{UV})^A = \SI{0.55+-0.09}{\kilo\parsec}$ for galaxy A and $\Delta(\lya-\text{UV})^B=\SI{3.7+-0.2}{\kilo\parsec}$ for galaxy B. Compared to the \citet{Claeyssens2022LlamasPaper1} sample, the offset of galaxy B ranks the largest at face value. However, when normalized to the size of the UV model (using the definition of ``elliptical distance,'' $\Delta_\text{ell}$, in Equation (3) of \citet{Claeyssens2022LlamasPaper1}), the offsets of A and B both qualify as ``internal spatial offsets'' for $\Delta_\text{ell}^A\approx\num{0.06}$ and $\Delta_\text{ell}^B\approx\num{0.02}$ are well below unity.

    \subsubsection{Spatial Distribution of Line Parameters}\label{sec:spatial}
    
    In this section we describe the procedure to extract and characterize the spectral line profile in resolved regions of the system. Following \citet{Claeyssens2019LyaSpectralVariations}, we started from our best source plane model of \lya~emission (see \autoref{sec:radial_profiles}) and evaluated it in a grid covering the delensed coordinates, at a pixel size of \ang{;;0.03}. The resulting image was then fed to the \textsc{vorbin} package \citep{Cappellari2003VorbinPaper}, which produces a tessellation of the source plane into regions of roughly equal flux. 
    
    With this method, we constructed two different tessellations: first, a high-resolution tessellation of 50 bins with a median S/N of $\sim20$ that we used for fitting the AG profiles and second, a lower-resolution tesselation of 24 spatial bins with a median S/N of $\sim30$, used to fit the galactic wind models (see \autoref{sec:flareon}), which require higher S/N for producing reliable results.  Then, for a given tessellation, we extracted the spectrum from each region by combining all  the image plane spaxels that trace back to it and coadded the corresponding continuum-subtracted spectra. We made sure that the traced image plane regions have at least four contiguous spaxels and they are larger than the PSF at least in one direction.
    
    We characterized the spectral properties of the redshifted \lya~line in each bin by fitting an AG profile (see \autoref{sec:asym-gauss}). An analysis of physically motivated models for the observed \lya~profiles is postponed to \autoref{sec:flareon}.
    
   Before fitting, each bin was assigned a systemic redshift based on its spatial overlap with UV emission of the galaxies. Bins having more than 50\% of their UV flux inside the continuum masks defined in \autoref{sec:integrated_spectrum} were assigned the systemic redshift of the corresponding galaxy (i.e., $z_A=\num{2.9257}$ for galaxy A and $z_B=\num{2.9238}$ for galaxy B). For all the other bins we set $z_\text{mean} = (z_A + z_B)/2 = 2.92475$.
    
    The fit is well behaved and the marginalized posterior distributions are not multimodal, except for a single bin of the 50-bin tessellation located at the outskirts of the halo whose spectrum does not have enough S/N for MCMC convergence. We used the integrated flux yielded by the fitted model to estimate both the intrinsic SB  of each bin (dividing by the bin solid angle) and the total luminosity of the system (dividing by bin magnification\footnote{Defined as the ratio between the total solid angle spanned by the image plane spaxels associated with that bin and its solid angle in the source plane.} and taking the sum). In this way, we obtained $L_{\lya}=\SI{6.2+-1.3e42}{\erg\per\second}$ for the whole \name~system\footnote{The value was corrected for a Galactic attenuation of 0.075 mag at \SI{4771}{\angstrom},  which was interpolated from the \citet{SchlaflyFinkbeinerMWReddening} tables. Also assumes a 20\% error due to magnification.} based on the 50-bin tessellation.
    
    \begin{figure*}[!htb]
        \centering
        \includegraphics[height=0.42\linewidth]{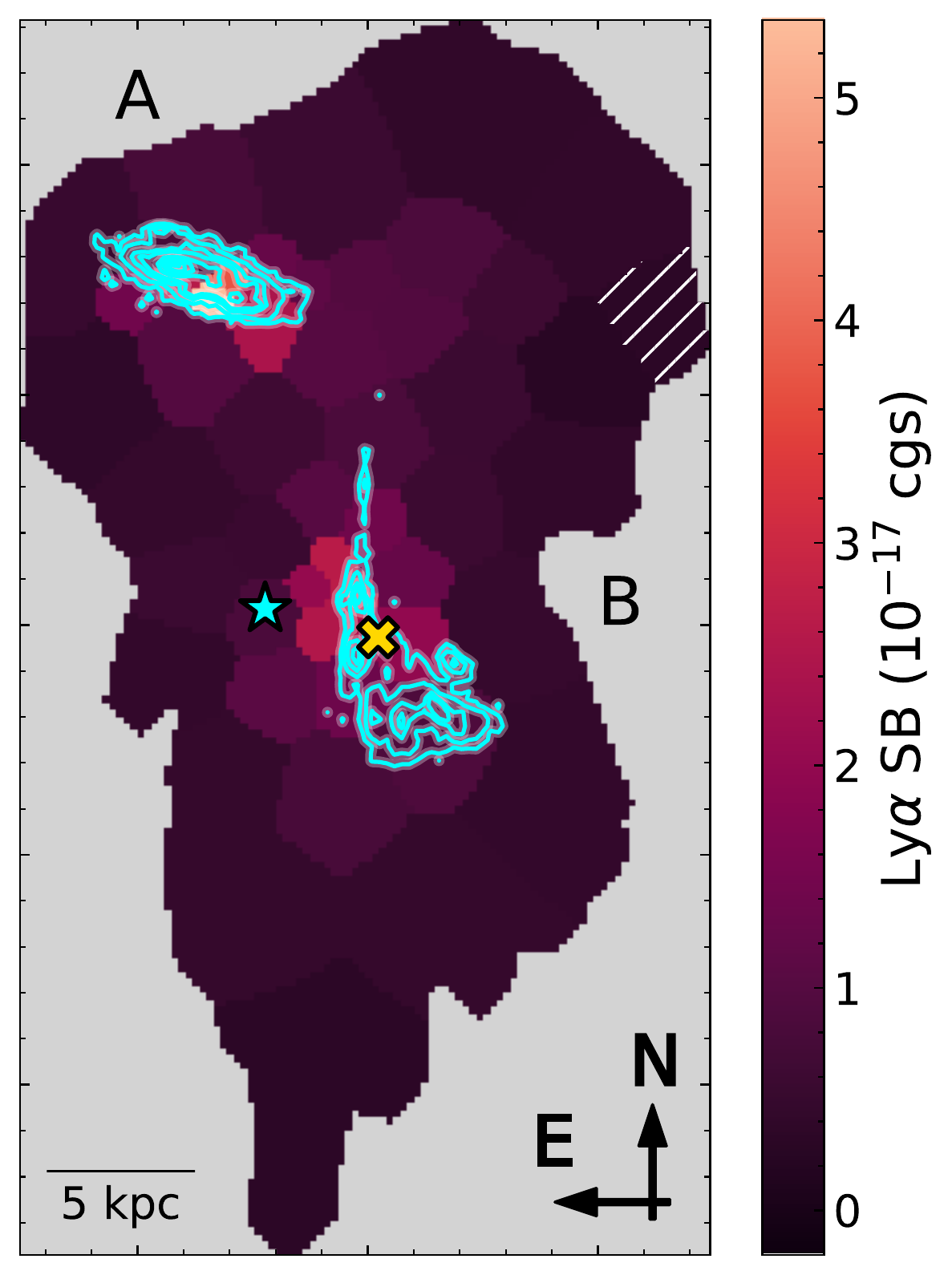}~
        \includegraphics[height=0.42\linewidth]{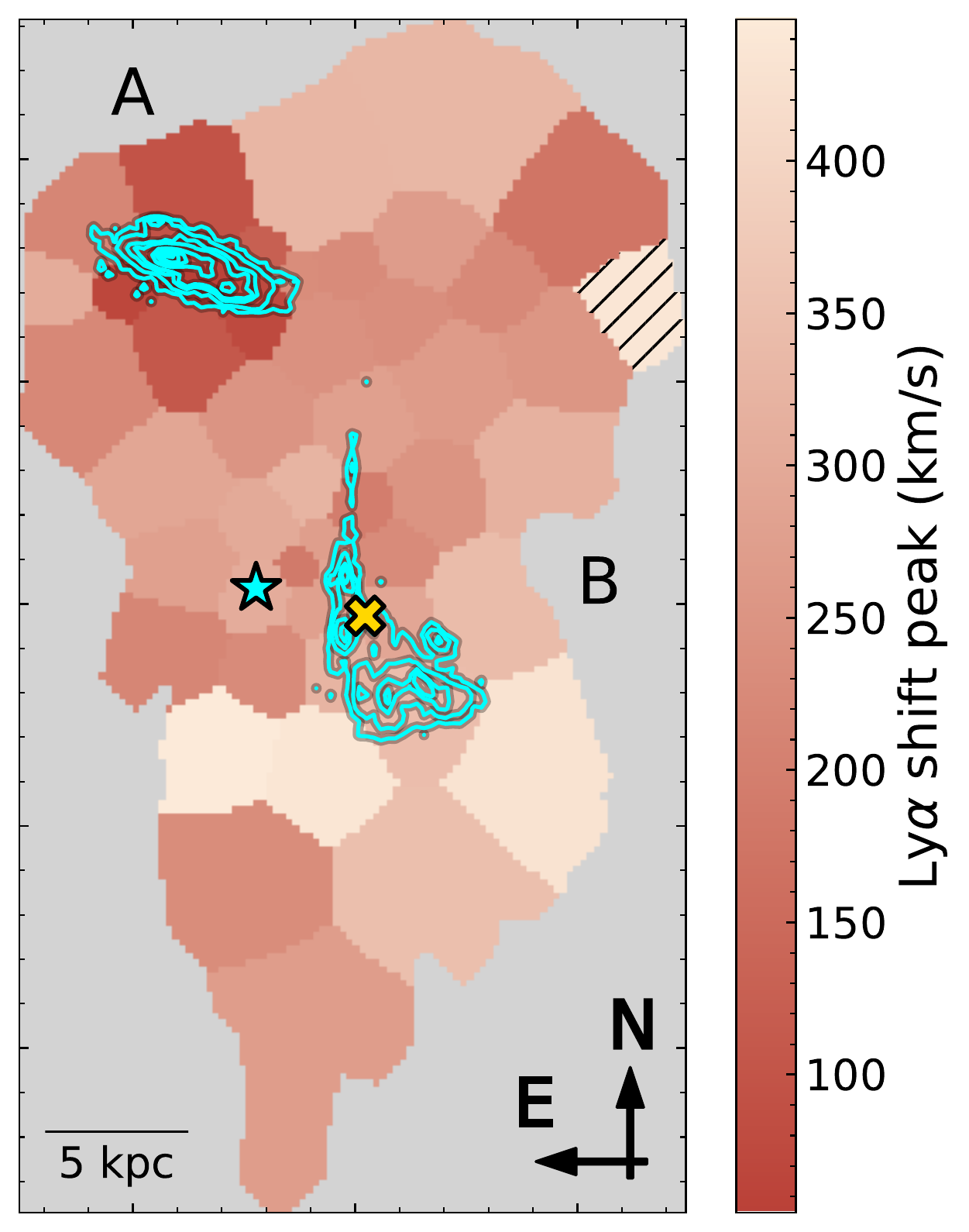}~
        \includegraphics[height=0.42\linewidth]{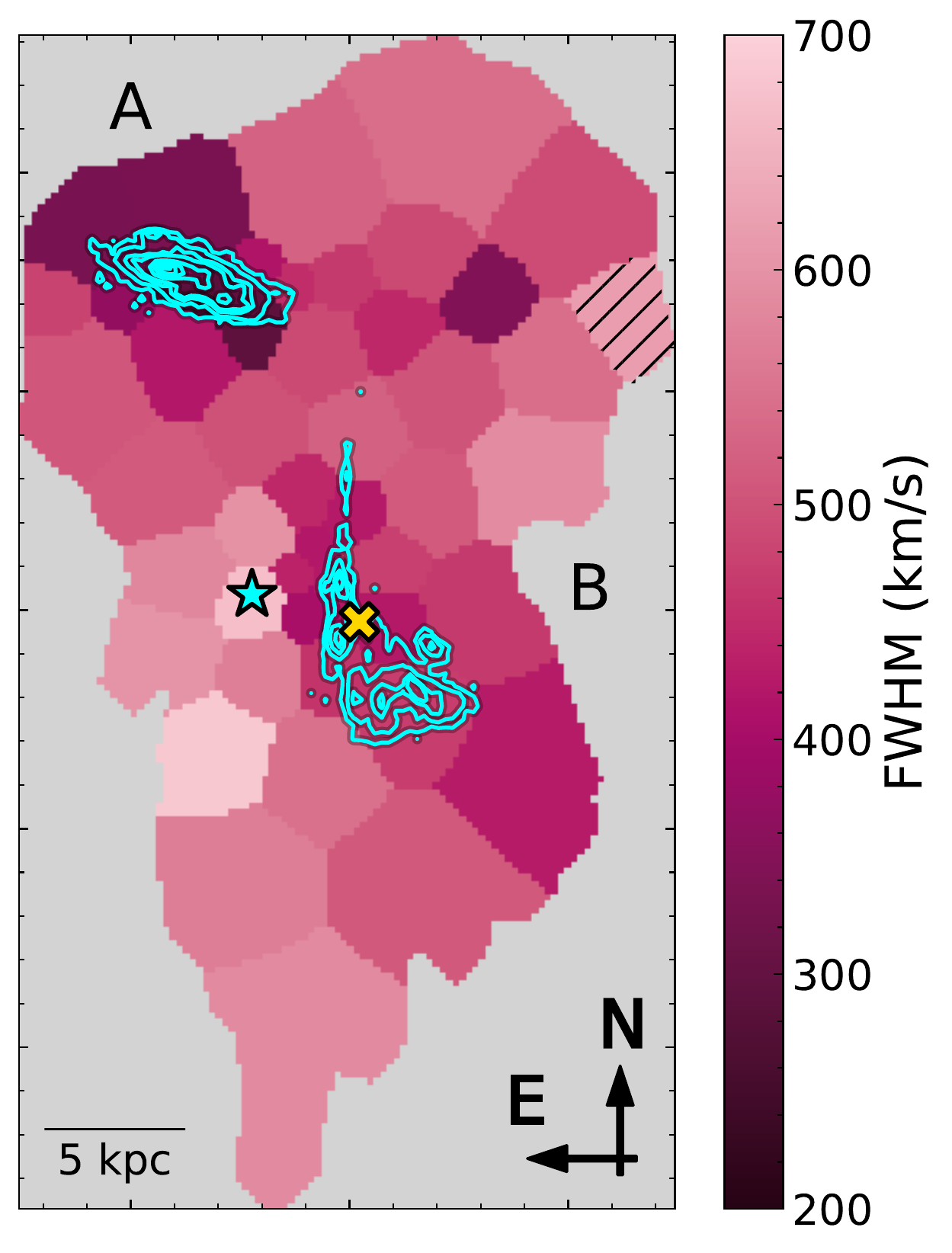}
        \caption{Source plane map of AG-fitted parameters for the 50-bin tessellation. In all panels, cyan contours indicate the locus of the UV continuum obtained from a pixelated source plane reconstruction of A.2 and B in the ACS F606W image, indicating the 3\%,  6\%, 13\%, 19\%, 38\%, 56\%, 75\%,  and 94\%  levels of the maximum (\SI{1.2e-20}{\erg\per\second\per\centi\meter\squared\per\angstrom}). The cross marker is the centroid of the dust continuum detected in ACA \citep{Solimano2021MolecularGasIntermediateMass}. The star symbol indicates the source plane position of \name-LAE. Also, we mark with a hatched pattern the bin that we rejected from the analysis due to poor fitting results. Left panel: \lya~SB. Middle panel: Velocity shift of the red peak relative to the adopted systemic redshift, either 2.9257, 2.9238, or 2.9247 depending on bin membership to A, B, or the diffuse halo, respectively (see \autoref{sec:spatial}). Right panel: FWHM of the \lya~line.} \label{fig:src_plane_maps}
    \end{figure*}
    
\begin{figure*}[!htb]
        \centering
        \includegraphics[width=\linewidth]{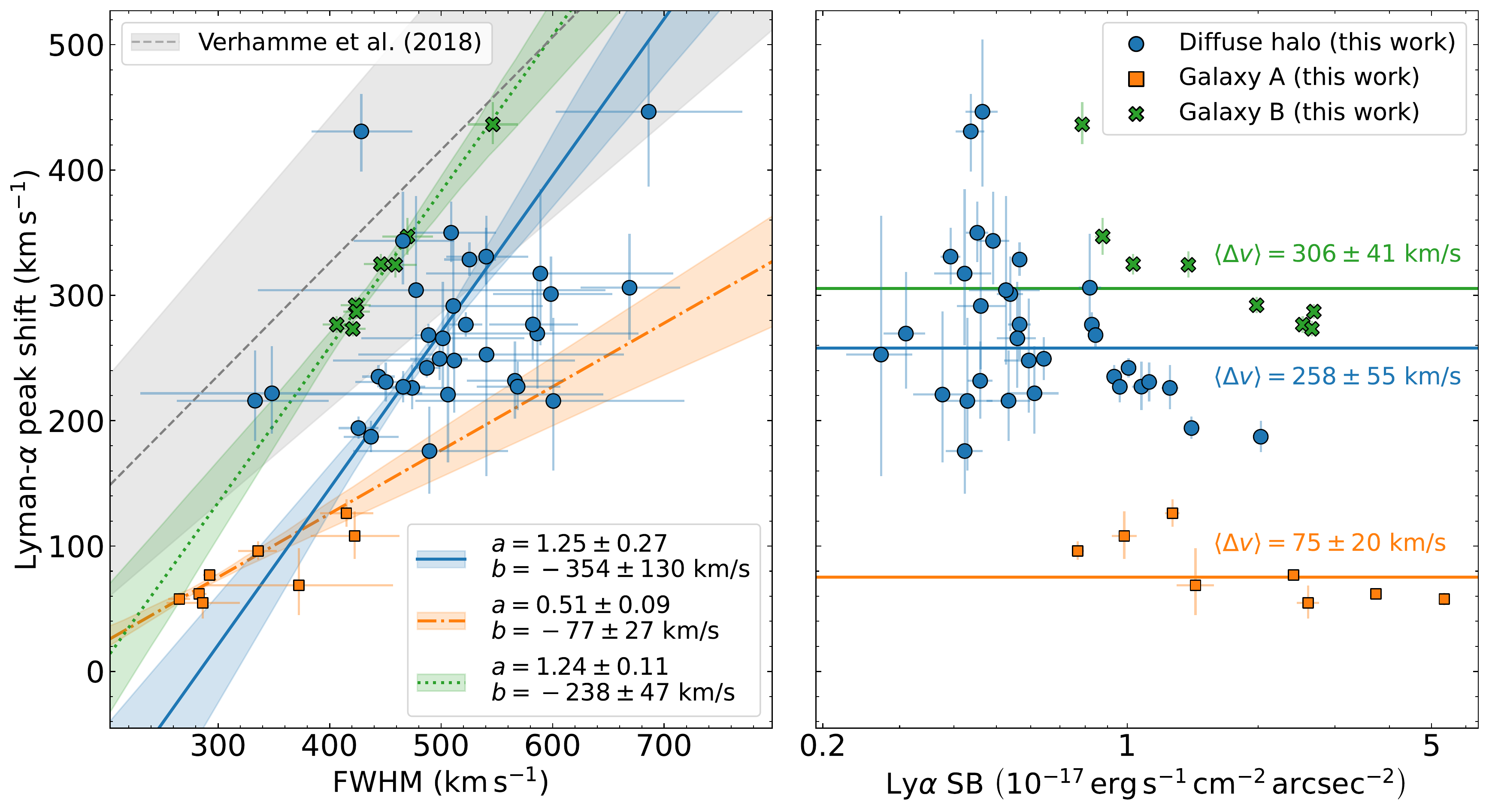}
        \caption{Left panel: Correlation between \lya~line FWHM and peak shift velocity for the individual regions of the 50-bin tessellation of \name, obtained through parametric modeling of the line as an AG  profile. Marker shapes and colors separate the three components of the \name~system, namely the cores of galaxies A (orange squares) and B (green crosses) and the halo diffuse emission (blue circles). For the latter, the systemic velocity was set to $z_\text{mean}=(z_A + z_B)/2=2.92475$. The best fitting straight line is also displayed for each of these components, along with their 68\% confidence intervals shown as a shaded area around the lines. The dashed gray line traces the empirical relation found by \citet{Verhamme2018SystemicRedshiftLya}. Right panel: \lya~peak shift velocity versus SB. Horizontal lines mark the uncertainty-weighted average peak shift velocity  $\left<\Delta v\right>$ for each of the three components.} 
        \label{fig:shift_vs_fwhm}
    \end{figure*}

    \autoref{fig:src_plane_maps} shows the source plane tessellation with the bins color-coded according to the fitted SB, peak shift velocity, and line width.
    The leftmost panel shows the source plane distribution of SB, which peaks near the position of galaxies A and B (cyan contours) and rapidly declines toward the outskirts. However, instead of being two distinct halos, the two resolved \lya~SB peaks are connected by a low-SB ``bridge'' or filament. The peak velocity  approximately ranges from \num{50} to \SI{450}{\kilo\meter\per\second} across the halo, as seen in the middle panel of \autoref{fig:src_plane_maps}. The lowest values are the ones associated with galaxy A, with an uncertainty-weighted mean and standard deviation of \num{75} and \SI{20}{\kilo\meter\per\second} respectively. The highest values are instead associated to either galaxy B or to bins at the outskirts of the halo.  The right panel also shows a large spread in line FWHM, spanning from \num{265} to \SI{690}{\kilo\meter\per\second}. We observe that the bins covering the UV continuum typically have samller widths than the bins of the diffuse halo, in agreement with the trends observed by \citet{Claeyssens2019LyaSpectralVariations} and \citet{Leclercq2020SpatiallyResolvedLyaHalosMHUDF}. Qualitatively, the presence of this pattern seems to confirm the results obtained above for the integrated apertures (see \autoref{sec:integrated_spectrum}). A remarkable exception, however, is the bin covering the faint companion \name-LAE (indicated with a cyan star in \autoref{fig:src_plane_maps}), which exhibits the second largest line width of the system ($\approx \SI{670}{\kilo\meter\per\second}$).  Similarities between the FWHM and peak shift maps should not come as a surprise. Recent studies have found that these two quantities are positively correlated, both among integrated measurements across different objects \citep{Verhamme2018SystemicRedshiftLya} and within resolved regions of individual halos \citep{Claeyssens2019LyaSpectralVariations, Leclercq2020SpatiallyResolvedLyaHalosMHUDF}. From a theoretical perspective, the relation is expected to arise as a natural consequence of resonant scattering, as confirmed by radiative transfer calculations in simple geometries \citep[e.g.,][]{Schaerer2011GridOfLyaRT, ZhengAndWallace2014AnisotropicLya, Song2020LyaRTSpectrumAndSB}. In contrast, the relation is not found in more complex, high-resolution simulations that include full radiative hydrodynamics \citep{Behrens2019LyaSimulatedGalaxiesEoR, Mitchell2021TracingSimulatedCGMLya}. In \autoref{fig:shift_vs_fwhm} we show the FWHM versus peak shift values for the 49 spatial bins with good fits in the 50-bin tessellation, in comparison with the \citet{Verhamme2018SystemicRedshiftLya} empirical relation. Our data also exhibits the FWHM-shift correlation with an uncertainty-weighted Pearson's coefficient of $r=0.76_{-0.21}^{+0.09}$ ($p<\num{e-5}$), where the errors have been estimated with bootstrapping. Motivated by this result, we performed orthogonal linear regressions using SciPy's \verb|odr| module on the data separated in the three systemic redshift groups. The best-fit lines are plotted in the left panel of \autoref{fig:shift_vs_fwhm} along with their corresponding 68\% confidence intervals. Each line is described by a slope $a$ and a zero-point $b$. We found that the slopes for the diffuse halo and galaxy B bins agree at $a\approx 1.25$ within a $1\sigma$ error, while their zero-points differ significantly by about \SI{120}{\kilo\meter\per\second}. These offsets put the resolved relations for B and the halo slightly below \citeauthor{Verhamme2018SystemicRedshiftLya}'s empirical relation ($a=0.9\pm0.14$; $b=-34\pm60$), but the slopes are still consistent within $1\sigma$.  Galaxy A's data, on the other hand, favors a much shallower relation ($a=0.51\pm0.09$), which cannot be brought into agreement with \citeauthor{Verhamme2018SystemicRedshiftLya}'s, but is within the broad range of slopes observed by \citet{Leclercq2020SpatiallyResolvedLyaHalosMHUDF} in resolved individual halos. Additionally, we investigated the effect of SB in the line profile. Just as in the two objects studied by \citet{Claeyssens2019LyaSpectralVariations}, we observe that above a certain SB threshold, brighter bins have smaller peak velocity shifts, as shown in the right panel of \autoref{fig:shift_vs_fwhm}.

    \subsubsection{FLaREON Models}\label{sec:flareon}
    Here, we analyze the binned spectra with physically motivated models. We used the publicly available FLaREON package \citep{GurungLopez2019FLaREON} to fit template spectra for every bin. The code performs a nonlinear search in a grid of \lya~line template profiles from precomputed Monte Carlo radiative transfer (RT) simulations from the \textsc{lyart} project \citep{Orsi2012GalacticOutflowsLyaEmitters}. The simulations only include two radiative transfer effects, namely resonant scattering and dust absorption. Hence, for the interpretation of these models one has to assume that all the \lya~photons are produced in the central galaxies and then scattered away from resonance in the CGM. Other mechanisms for the production of extended \lya~emission, such as fluorescence and gravitational cooling, are discussed in \autoref{sec:discussion}.  
    
    Three geometries are currently available within FLaREON: The first two are a spherical, expanding thin shell of isothermal gas around a point source of \lya~photons (hereafter TS) and a galactic wind geometry (GW), which is identical to the TS geometry except for the density distribution of the gas. While the TS assumes all the gas is concentrated in a thin shell at a fixed distance of the source, in the GW geometry the source lies in an empty spherical cavity surrounded by a spherical distribution of isothermal gas with radially declining gas density. Finally, FLaREON also offers a biconical outflow geometry, which combines a static uniform medium with a bicone of lower-density, expanding gas. All geometries are governed by the same three parameters: the expansion velocity $V_\text{exp}$, the neutral hydrogen column density $\log{N_\text{H I}}$ and the pure absorption (dust) optical depth $\log{\tau}$.
    
    While these models were originally developed to describe the profiles of spatially unresolved spectra of LAEs, here we use them to model individual bins of our source plane tessellation. The main caveat of this approach is that the emission source will not be, in most cases, at the center of the binned region. This means that a typical \lya~photon produced by the central galaxies will need to travel a larger distance (thus with a higher probability of absorption or scattering) to escape from any given bin than what the models imply. However, \citet{Chen2021LensedLyaSuperWinds} proposed that if the \lya~photons do not originate inside the cloud, the problem can still be described with just half of the expanding sphere. Following their argument, the isotropy of the TS and GW geometries implies that a  signal arising from a single hemisphere would have the same line profile as the full spectrum but with a reduction in amplitude. This property suggests that one can approximate the full halo as a collection of half-expanding clouds, where now the expansion velocity is measured relative to a reference point inside each cloud along the line of sight. Since the biconical wind geometry is not isotropic, we did not try that geometry in our modeling.

    The code requires the input of a systemic redshift in order to transform observed wavelengths into rest-frame wavelengths. However, the choice of systemic redshift is known to have a significant influence on other parameters \citep{GurungLopez2019FLaREON}. Due to this, one could in principle infer the systemic redshift from the \lya~profile shape alone. To explore this idea, we started by fitting FLaREON models with $z_\text{sys}$ as a free parameter to the integrated spectra of galaxies A and B (see \autoref{sec:integrated_spectrum}), because their systemic redshifts are known and their S/N is the highest. We ran an MCMC fitting scheme similar to the one used for fitting the AG for the two isotropic geometries, TS and GW. The total number of free parameters for each model is five, since we fit an amplitude scale factor in addition to the three main parameters and the systemic redshift. 
    
    \begin{figure*}[!htb]
        \centering
        \includegraphics[width=\linewidth]{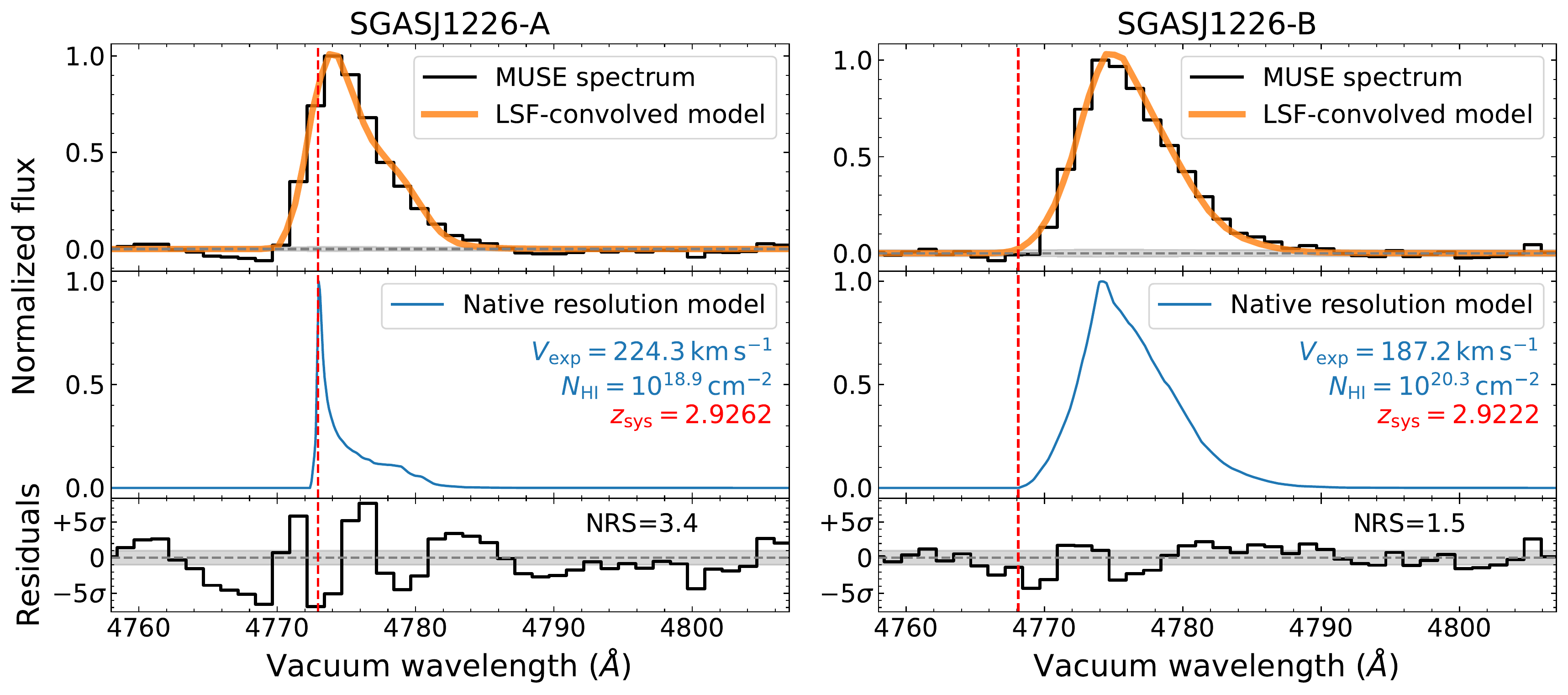}
        \caption{Best-fit FLaREON GW models for the integrated spectra of galaxy A (left) and galaxy B (right). The top row of panels shows the MUSE data (black step line) along with the best-fitting FLaREON model (thick orange line) convolved with the LSF. The middle row shows the same models prior to LSF convolution (blue line). The parameters for the profiles are indicated in the inset text. The residuals are shown in the bottom row, with the ``normalized residual scatter'' (NRS), i.e., the standard deviation of the residuals weighted by the error spectrum. The gray filled area indicates the $\pm1\sigma$ error.}
        \label{fig:flareon_gw_profiles_AB}
    \end{figure*}
    
    \autoref{fig:flareon_gw_profiles_AB} shows the best-fit GW models and residuals for the integrated spectra of A and B in the left and right panels, respectively. Remarkably, the systemic redshift order ($z_A > z_B$) is correctly inferred by the models, despite the uniform priors and the fact that the line centroid of B is actually redder than A's. Moreover, the fitted redshifts $z_A = 2.9262 \pm 0.0001$ and $z_B = 2.9222 \pm 0.0001$ are only \SI{50+-10}{\kilo\meter\per\second} and \SI{-163+-10}{\kilo\meter\per\second} from the nebular redshift solution of A and B, respectively, consistent with the scatter of the different solutions (\autoref{sec:sed}). The inferred expansion velocities and neutral column densities are $V_\text{exp}^A = \SI{224.3}{\kilo\meter\per\second}$ and $V_\text{exp}^B = \SI{187.2}{\kilo\meter\per\second}$; and $\log(N_\text{H I}^A/\si{\per\centi\meter\squared}) = 18.9$ and $\log(N_\text{H I}^B/\si{\per\centi\meter\squared}) = 20.3$. These two crucial parameters are constrained by the data, whereas the optical depth is not.

    The best-fit velocity for galaxy A is higher than the best value for galaxy B. This already gives an interpretation to the line profile being narrower in A: since the medium is moving at higher velocities, the \lya~photons need to experience on average fewer scattering events to shift their frequency out of resonance, because the required shift is smaller. This effect is coupled with the lower column density of neutral hydrogen in A with respect to B. A lower density of neutral hydrogen atoms reduces the number of scattering events that a photon will experience before escaping. At a fixed Doppler shift and gas velocity, a \lya~photon has higher chances of escaping if there are fewer atoms along its path.
    
    We also fitted the diffuse-only spectrum in this fashion, that is, by having the systemic redshift as a free parameter. Interestingly, the best-fit redshift is $z_\text{sys}=2.9222$, extremely close to the best value given to B's model. The column density inferred by the model is also similar to B's ($N_\text{H I}^\text{diffuse} \approx 10^{20.4}\,\si{\per\centi\meter\squared}$), but the expansion velocity of \SI{198.3}{\kilo\meter\per\second} is in between $V_\text{exp}^A$ and $V_\text{exp}^B$. These results are, however, more difficult to interpret since we have explicitly excluded the MUSE spaxels with UV continuum emission (a proxy for a high SFR), so there is no source of \lya~photons along the line of sight under the assumption of a pure scattering scenario. In this sense, the high column density should be regarded as an average integrated along random optical paths of the escaping photons, rather than along the line of sight between the observer and the source. This is the meaning that we will give to our subsequent results on the binned regions' spectra.
    
    The models using the TS geometry yield similar results, but the $\chi^2$ is larger in all the three fits and even though the $z_A > z_B$ property is recovered, the predicted redshifts are less accurate. For this reason, plus the fact that the line shape does not vary dramatically across the halo at the MUSE resolution, we adopted the GW as our fiducial geometry to model the spectra from the individual bins. 
    
\begin{figure*}[!hbt]
     \centering
    \includegraphics[width=0.3\linewidth]{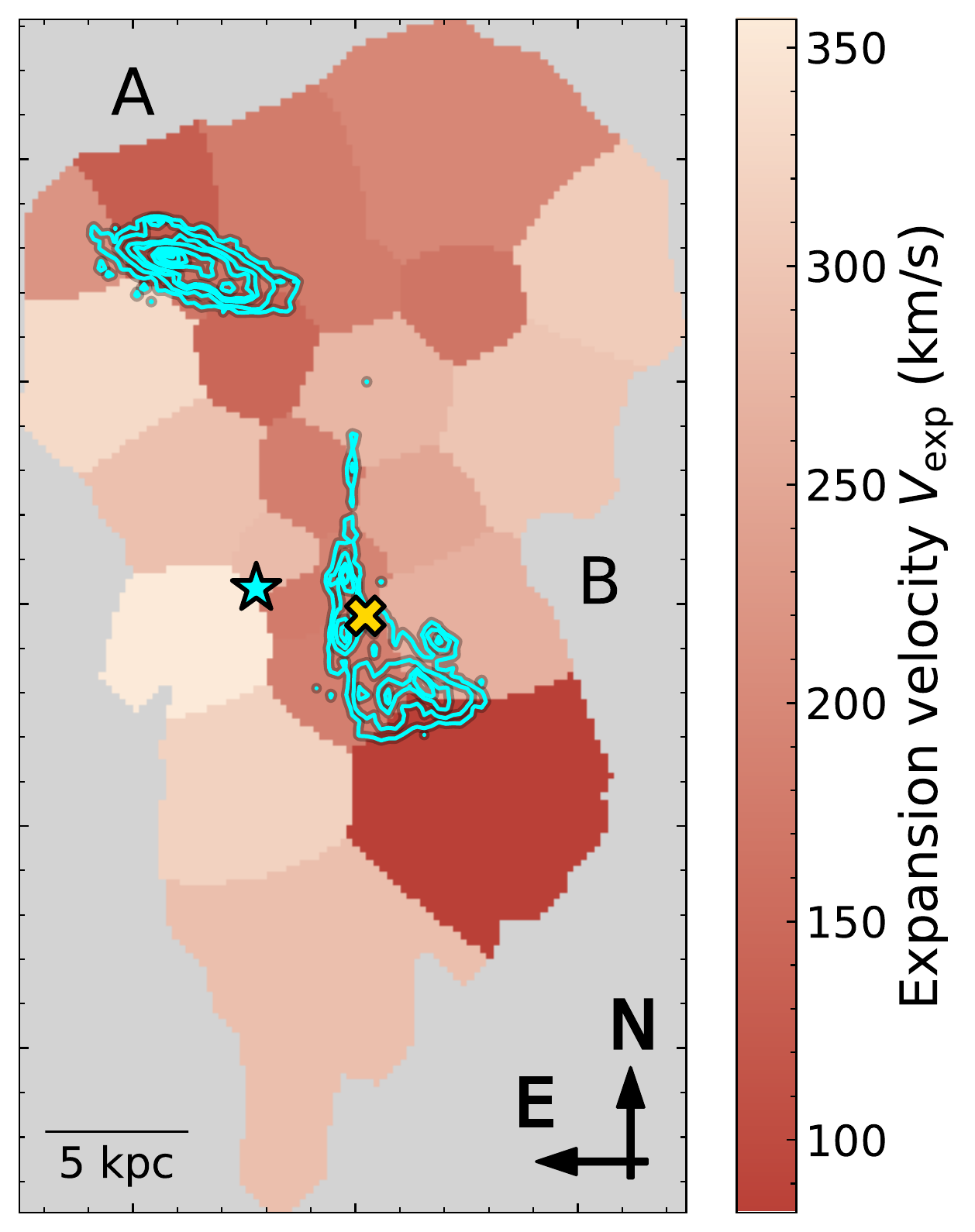}~
    \includegraphics[width=0.3\linewidth]{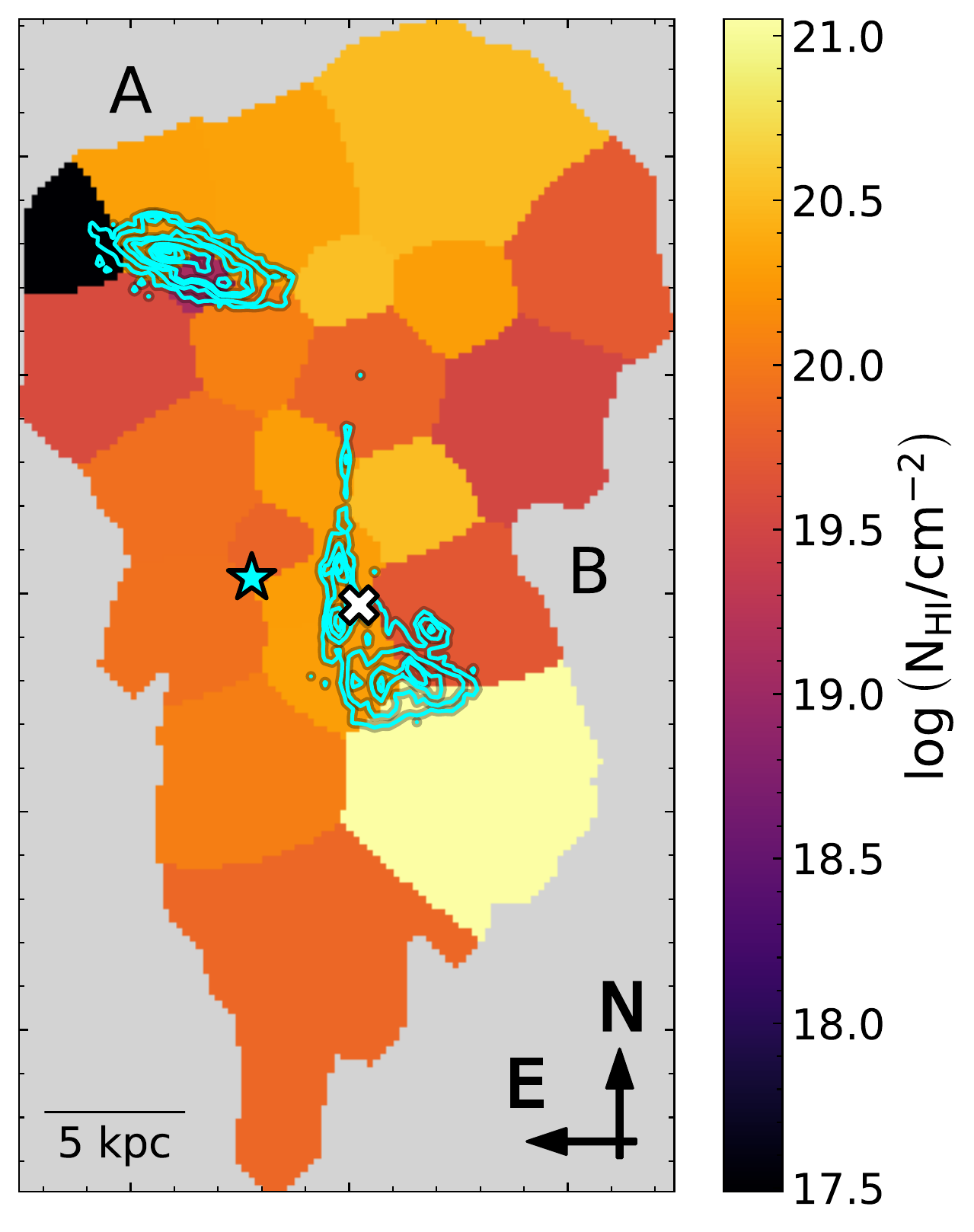}~
    \includegraphics[width=0.3\linewidth]{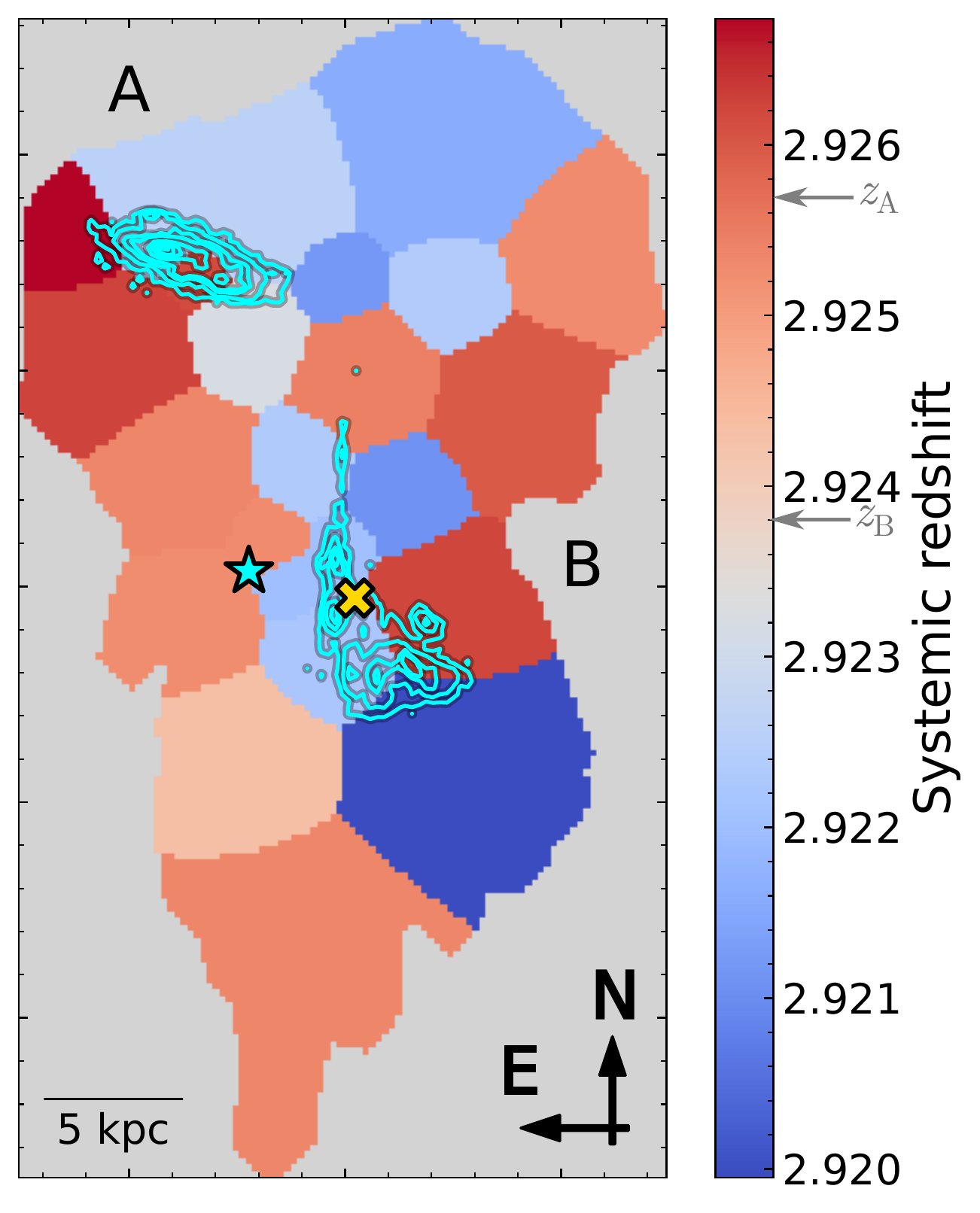}
     \caption{Source plane maps of FLaREON GW fitted parameters. The contours and markers are the same as those in \autoref{fig:src_plane_maps}. Left panel: Wind expansion velocity. Middle panel: Neutral hydrogen column density. Right panel: Systemic redshift inferred from the \lya~line. Nebular redshift solutions for A and B are indicated in the colormap axis.}
     \label{fig:gw_sp_24}
    \end{figure*}

    Finally, given that our FLaREON GW model recovered the systemic redshifts reasonably well and given our lack of systemic redshifts for most of the binned regions, we also let that parameter vary freely in the fitting procedure for the binned regions. Here we used the 24-bin tessellation, since we required a higher S/N per spectrum. The source plane maps of the best-fit GW parameters are shown in \autoref{fig:gw_sp_24}. Again, the dust absorption optical depth is not constrained by the data, and hence we are unable to make inferences on this parameter.

    The expansion velocities inferred by FLaREON under the GW geometry range from \num{80} to \SI{360}{\kilo\meter\per\second}, with an uncertainty-weighted average of $\left<V_\text{exp}\right> = \SI{211+-3}{\kilo\meter\per\second}$ and a standard deviation of $\SI{62.5+-3}{\kilo\meter\per\second}$. Interestingly, \autoref{fig:gw_sp_24} suggests some degree of spatial correlation between the expansion velocity values across the halo: Low-velocity bins are clustered along the north-south axis (vertical in the figure, and hereafter referred to as the ``major axis'' of the halo), including bins associated with the continuum of the central galaxies. Similarly, high-velocity bins are associated with diffuse-only regions at the outskirts of the halo, and they are found on both sides of the major axis and are also clustered in the same direction. These results suggest that on halo scales the medium is not expanding isotropically, but rather with a preferred direction. In other words, they might indicate that the gas is traveling faster in a direction perpendicular to the line connecting the two galaxies (e.g., due to decreased resistance from the environment). 
    
    In terms of column density, the models are distributed around $\log\left(N_\text{H I}/\si{\per\centi\meter\squared}\right) = 20$, with a $3\sigma$-clipped standard deviation of \num{0.4} dex. After discarding three outlier bins ($\Delta\log{N_\text{H I}} \gtrsim 1$ dex),  we found that the path-integrated column density is remarkably uniform across the halo. Under the interpretation proposed above, this would mean that the actual column density declines as a function of the radial distance to the \lya~sources.
    
    Finally, in the right panel of \autoref{fig:flareon_gw_profiles_AB} we plot the systemic redshifts predicted by FLaREON for each bin. If we interpret them as the zero velocity of the reference point inside of each GW model, its complex spatial structure and a range of values exceeding differences of \SI{700}{\kilo\meter\per\second} would indicate that the outflow is modulated by complex underlying kinematics. Such complexity would naturally arise in a scenario where the two main galaxies are subject to dynamical interactions, for example, those observed in a merger. Further discussion of the potential impact of the interaction between A and B is made in \autoref{sec:kinematics}.

   \subsection{Low-Ionization Absorption Lines}\label{sec:absorption}
    Independent insight into the kinematics of the system can be obtained from the study of absorption features in the spectrum, although they only probe intervening gas along the line of sight to the central galaxies. As mentioned in \autoref{sec:sed}, previous studies of \name~have obtained redshift solutions based on interstellar absorption lines that are $\sim \SI{200}{\kilo\meter\per\second}$ bluer than the nebular emission based solutions \citep{Koester2010ArcJ1226Discovery, Wuyts2012StellarPopsLensed}. 
    This result was later secured using the higher-resolution Magellan Echellette (MagE) spectrum of arc A.1 obtained as part of the MegaSaura Survey \citep{Rigby2018MegaSauraI}. Moreover, \citet{Gazagnes2018NeutralGasPropertiesLyC} fitted Voigt profiles to the Lyman-series absorption lines (from Ly$\beta$ \SI{1025.7}{\angstrom} to Ly6 \SI{930.8}{\angstrom}), obtaining a central velocity of \SI[multi-part-units=single]{-264\pm21}{\kilo\meter\per\second} (\SI{-218}{\kilo\meter\per\second} if we convert to our nebular redshift solution).
    
\begin{figure}
        \centering
        \includegraphics[width=\columnwidth]{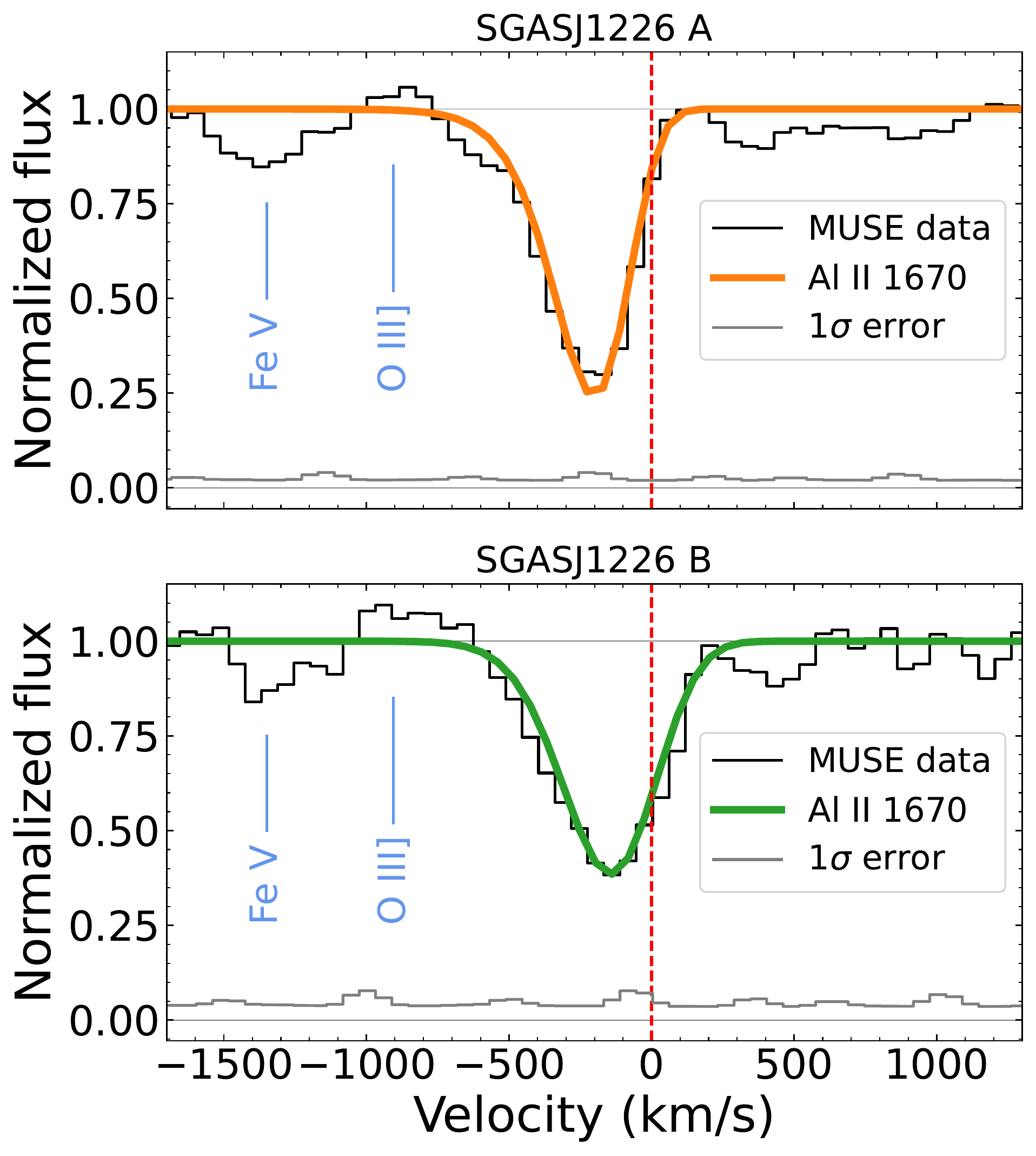}
        \caption{Spatially integrated \ion{Al}{2}$~\lambda 1670$ absorption spectra of galaxies A (top panel) and B (bottom panel). The solid colored lines are the best-fitting AG profiles, while the vertical dashed line indicates the systemic velocity of each galaxy. The photospheric \ion{Fe}{5} line and the nebular \ion{O}{3}] line are also annotated in the figure.}
        \label{fig:absorption}
    \end{figure}

    Here, we complement these results by measuring the absorption velocity in our MUSE data, including galaxy B. But since the Lyman-series lines lie below the MUSE wavelength cutoff, we opted for the low-ionization \ion{Al}{2}$~\lambda 1670$ line as an alternative tracer. We extracted spatially integrated spectra from the global apertures for A and B defined in \autoref{sec:integrated_spectrum}. Both spectra show a very similar profile that includes an asymmetric blue tail, and thus were modeled with an absorption AG profile (see \autoref{fig:absorption} and \autoref{sec:asym-gauss}). We found $v_A = \SI{-197+-4}{\kilo\meter\per\second}$ and $v_B = \SI{-142+-13}{\kilo\meter\per\second}$ . Also, despite the line being narrower in A than in B ($\text{FWHM}_A = \SI[multi-part-units=single]{300+-7}{\kilo\meter\per\second}$ versus $\text{FWHM}_B = \SI[multi-part-units=single]{373+-15}{\kilo\meter\per\second}$), their EWs are consistent with each other, for $\text{EW}_0^A = \SI{5.8+-0.5}{\angstrom}$ and $\text{EW}_0^B = \SI{5.6+-0.2}{\angstrom}$.

    The combination of blueshifted absorption lines and redshifted \lya~emission is very common among $z\sim3$ LBGs \citep[e.g.,][]{Steidel2003LBGsAtRedshift3}, and it is often interpreted as a telltale signature of (spherical) galactic outflows \citep{Verhamme2006LyaRT3DPaperI}. 
    
   \section{Discussion}\label{sec:discussion}
   
   \subsection{Kinematics of \name}\label{sec:kinematics} 
   A major goal of this paper is to understand the physical configuration of the \name~system, by building a picture that includes both its spatial and kinematic properties. Firstly, the lens model reveals that the UV centroids of the two main galaxies of the system are separated only by \SI[multi-part-units=single]{14.3+-1.4}{\kilo\parsec} in projection (see \autoref{sec:radial_profiles}). This fact, together with the small systemic velocity offset between the two galaxies ($\Delta v \approx \SI{145}{\kilo\meter\per\second}$), suggests that they are gravitationally interacting. The interaction hypothesis is further motivated by the reconstructed source plane  continuum morphology (contours in \autoref{fig:src_plane_maps}), which exhibits a very distorted galaxy B resembling the tidal tails or collisional rings seen in galaxy interactions at low redshift \citep[e.g., ][]{Darg2010GalaxyZooMergers}. The source plane morphology, however, should be interpreted with care, because the distorted appearance of galaxy B can also be due to (1) strong dust attenuation in some parts of the ISM that conceal the full shape of the galaxy, or to (2), geometrical artifacts that arise from the lens modeling. With the available data, we cannot distinguish whether the galaxies are experiencing a single visit flyby or the early stage of a major merger. On one hand, the lack of an ongoing starburst in both galaxies \citep[their specific SFRs are consistent with the main sequence,][]{Solimano2021MolecularGasIntermediateMass} indicates that the interaction (if any) is not currently boosting the SFR, as it would be expected in some stages of a major merger \citep[e.g.,][]{Hopkins2008CosmoFramework1Mergers}. On the other hand, the extended \lya~emission implies a significant amount of gas between and around the pair of galaxies, thus favoring a merger scenario in which the interaction was able to strip gas away from the ISM and push it into the CGM \citep[see, for example, ][]{Yajima2013ExtendedLyaFromInteractingGalaxies}. Also, we observed in \autoref{sec:flareon} that the fitted systemic redshift map exhibits a complex morphology and a large range of values, which provide hints of complex underlying kinematics similar to the expectation for galaxy mergers \citep[e.g.,][]{Sparre2022GasFlowsInGalaxyMergers}. But in order to confirm or reject the merger activity as a driver of the gas motions, it would be necessary to obtain resolved spectroscopy of a nonresonant line, such as H$\alpha$ or $[\textsc{C II}]\,\SI{158}{\micro\meter}$, to trace the internal kinematics of the ISM where the effects of a merger would be more evident.
    
    The global kinematics of the halo are clearly dominated by outflow motions, as revealed by the characteristic redshifted and red asymmetric \lya~profile. Under simple isotropic outflow geometries such as the TS and GW considered here, RT calculations always predict an enhanced red peak with a broadened red tail starting from relatively low expansion velocities. Extra evidence for outflows is seen in the absorption signature of low-ionization metal lines (see \autoref{sec:absorption}), which are thought to trace the same gas phase as \lya. The signature is seen  toward the UV continuum of both galaxies, A and B, but it remains unclear whether the outflows are launched independently by each galaxy or the outflow originates preferentially in one galaxy and the absorption appears in the second galaxy by a projection effect. The first case seems to be more likely, since the two galaxies have similar stellar masses and SFRs. Alternatively, the spatially coherent outflow signature could be linked to a large stream of receding gas stripped from the galaxies by the interactions described above.   

    Finally, 
    we did not find any evidence of CGM-scale rotation, although such a signal would be severely smeared by the RT effects. In fact, rotation-like gradients in the \lya~peak shift velocity of LAHs are very rare, with only one candidate out of six in the sample presented by \citet{Leclercq2020SpatiallyResolvedLyaHalosMHUDF}, the only one among the few resolved LAHs in the literature \citep{Claeyssens2019LyaSpectralVariations, Chen2021LensedLyaSuperWinds}. 
   
   \subsection{Powering mechanism}\label{sec:power}
    We have thus far assumed that the extended \lya~emission of \name~is exclusively explained by resonant scattering of photons produced in the central galaxies. In this section we explore alternative mechanisms that could also drive the observed properties of \name.
    
    \subsubsection{Gravitational cooling}\label{sec:grav_cooling}

    Galaxies need to accrete gas from the environment in order to sustain their growth over gigayear timescales \citep[e.g.,][]{Bouche2010ColdGasAccretion, Dave2012EquilibriumModelAnalytical, Scoville2017EvolutionISM, Tacconi2018PHIBSS2MainSequence, Walter2020EvolutionOfBaryons}. Simulations show that gas accretion can occur, for example, through cold streams of pristine gas from the IGM \citep[e.g., ][]{Dekel2009ColdModeAccretion, LHuillier2012AccretionVsMergers} or through inflows of previously ejected, metal-enriched gas in the CGM \citep{Springel2003CosmologicalHydroStarFormation, Oppenheimmer2010RecycledWindAccretion, Brook2014MaGICCBaryonCycle, Angles-Alcazar2017BaryonCycleInFIRE}. In either case, collisional interactions in the gas are expected to transform its gravitational binding energy into \lya~radiation. This mechanism is known as ``gravitational cooling,'' and it may contribute significantly to the development of extended LAHs \citep[e.g.,][]{Haiman2000LymanCoolingRadiation, Furlanetto2005LyaEmissionStructureFormation}. In this scenario, \lya~photons are created \textit{in situ}, and thus their peak shift should trace the line-of-sight velocity of the gas. Then, if there is inflowing gas between the observer and the center of the halo, the \lya~spectrum emitted from it will have a strong blueshifted peak and a blue tail \citep[e.g.,][]{Haiman2000LymanCoolingRadiation, Dijkstra2006LyaProtoGalaxiesII}. Such profiles can also be produced by scattering-only scenarios \citep[on a collapsing sphere of gas: e.g.,][]{Verhamme2006LyaRT3DPaperI} provided that the intervening gas has a velocity in the opposite direction of the propagation of the \lya~photons. In other words, the presence of an enhanced blue peak is a signature of inflowing motion, but not necessarily of gravitational cooling.   
     
     In \name, neither the integrated (\autoref{sec:integrated_spectrum}) nor the resolved spectra (\autoref{sec:spatial}) exhibit significant emission blueward of the systemic velocity. Only a few tentative blue peaks can be seen in the \lya~profiles of some halo regions (see regions 0, 2, and 5 in Fig. \ref{fig:profile_grid}). The lack of blueshifted emission cannot be entirely explained by intervening absorption in the IGM, because at $z=2.9$ the average IGM transmission at \SI{-200}{\kilo\meter\per\second} from the \lya~rest-frame wavelength is approximately 85\% \citep{Laursen2011IGMTransmissionLya}, high enough for a blue peak to be detectable.  The caveat is that this particular line of sight can have a higher-than-average neutral gas column density. We therefore conclude that the presence of significant inflowing motion along the line of sight is unlikely but not completely ruled out.
    
    \subsubsection{Fluorescence}\label{sec:fluoresence}
    Cool hydrogen gas in the CGM can be momentarily ionized after being exposed to Lyman-continuum radiation escaping the inner parts of the galaxy (from either an AGN or a  starburst region) or coming from the cosmic UV background (UVB). If the density is high enough, the atoms rapidly recombine and fall to the ground state, where a \lya~photon is emitted in a process called fluorescence. For this mechanism to produce extended \lya~emission, the gas needs to be distributed in high-density clumps with a low covering fraction, so the radiative transfer occurs preferentially in the surfaces of the clumps, resulting in a reduced number of scattering and absorption events with respect to the case of a homogeneous medium. In the approximation where scattering and dust absorption are negligible, this mechanism generates an intrinsic \lya~luminosity proportional to the production rate of ionizing photons. Here, we follow \citet{Valentino2016GiantLyaNebula} to estimate the ionizing photon rate $Q$ required to power the emission under the assumption that fluorescence is the only mechanism at play. From an observed, delensed $f_\text{esc}^{\lya}L_{\lya} = \SI{6.2+-0.2e+42}{\erg\per\second}$ we obtained
    \begin{equation}
        Q = \frac{L_{\lya}}{h \nu_{\lya} \eta_{\lya}} \approx \frac{5.6 \times 10^{53}}{f_\text{esc}^{\lya}}\,\si{\per\second},
    \end{equation}
    where $\eta_{\lya}=0.68$ is the fraction of ionizing photons converted into \lya~\citep{Spitzer1978PhysicalProcessesInterstellarMedium} and $h\nu_{\lya}=\SI{10.19}{\electronvolt}$ is the energy of a single \lya~photon. Since we cannot directly measure the production rate of ionizing photons (due to extreme ISM opacity at $\lambda_\text{rest} < \SI{912}{\angstrom}$) we need to rely on a longer-wavelength proxy such as the far UV luminosity ($\lambda_\text{rest}=\SI{1500}{\angstrom}$). But for a given $L_{\SI{1500}{\angstrom}}$ the actual production rate of ionizing photons $Q$ depends on the properties of the stellar population, particularly on the luminosity-weighted age and metallicity \citep{Steidel2001LyCfromLBGs, Smith2002IonizingContinuaMetallicity}. Fortunately, the stellar population synthesis analysis presented by \citet{Chisholm2019MegaSauraMetallicity} provides estimates of the ionizing photon production efficiency, $\xi_\text{ion} \equiv Q / L_{\SI{1500}{\angstrom}}$, for all galaxies in the MegaSaura sample including \name's arc A.1. Therefore, assuming that $\xi_\text{ion}$ is uniform across galaxy A and is the same for galaxy B, we can solve for $Q$ multiplying $\xi_\text{ion}$ by the total reddening-corrected UV luminosity of the system. For consistency with \citet{Chisholm2019MegaSauraMetallicity} we applied \citeauthor{Reddy2016FarUVAttCurve}'s \citeyearpar{Reddy2016FarUVAttCurve} attenuation law with \citeauthor{Chisholm2019MegaSauraMetallicity}'s best-fit color excess $E(B-V) = 0.13$ to the total demagnified luminosity inferred from ACS F606W photometry. After dereddening we obtained a UV luminosity of
    \begin{equation*}
        L_{1500}^{A+B} = \SI{3.4+-0.5e41}{\erg\per\second\per\angstrom}.
    \end{equation*}
    
    In Table 5 of \citet{Chisholm2019MegaSauraMetallicity}, the best-fit Starburst99 model for A.1's MagE spectrum implies a photon production efficiency of $\log{\xi_\text{ion}}=\num{12.74+-0.16}$ while the best-fit BPASS model favors a slightly higher value of $\log{\xi_\text{ion}}=\num{13.04+-0.16}$. Then, the total rate of production of ionizing photons by the galaxies in \name~is $Q=\SI{1.9+-0.7e54}{\per\second}$ for the Starburst99 model and $Q=\SI{3.7+-1.5e54}{\per\second}$ for the BPASS model. Taken at face value, these results imply that photoionization from young stellar populations would only account for $20\%-60\%$ of the total photon rate required to power the \lya~luminosity if we assumed $f_\text{esc}^{\lya}=0.082$, the value predicted by the \citet{Sobral2019LyaEscapeSFR} empirical relation based on the observed EW. Conversely, a  larger escape fraction (between \num{0.15} and \num{0.30}) would be needed to match the photon rates. 
    Now, these calculations only considered young stars as the  source of ionizing photons, but we cannot rule out the presence of an AGN that is obscured along the line of sight with our current data. 

    Also, we did not expect a significant contribution of metagalactic ionizing photons from the cosmic UVB, since the latest observational constraints imply that the UVB produces \lya~profiles at $z\approx3$ with peaks at the \SI{2e-20}{\erg\per\second\per\centi\meter\squared\per\angstrom\per\arcsec\squared} SB level \citep{Gallego2021ConstrainingUVBwithMUSE}, at least 3 orders of magnitude fainter than the observed SB peak of SGASJ1226 on halo scales---although its contribution to the profile can become relevant at large radii, in the interface with the IGM.
    
    In any case, we warn the reader that these ionizing photon budget considerations cannot constrain the contribution of fluorescence in the absence of an independent measure of the escape fraction and evidence of CGM clumpiness. The most direct test for fluorescence as a major contributor to the extended \lya~emission is the concomitant presence of extended H$\alpha$ emission \citep{MasRibas2017ExtendedLyaSatellites}. This is because under case B recombination \citep{OsterbrockAndFerland2006AstrophysicsNebulae}, the \lya~emissivity of the gas is 8.7 times the H$\alpha$ emissivity. In other words, the same regions of photoionized gas should glow in H$\alpha$ by a proportional amount, simultaneously solving the \lya~escape fraction and the question of \textit{in situ} \lya~production. Promisingly, \name~has been selected as a NIRSpec IFU target for the JWST Early Release Science ``TEMPLATES'' program \citep{Rigby2017TemplatesJWSTproposal} and thus resolved H$\alpha$ observations of the arc A.1 will become available. While the FoV of the instrument will not cover the whole LAH, it will certainly tell us if the H$\alpha$ extends beyond the ISM.

    \subsubsection{Emission from satellites and the nature of \arc-LAE}\label{sec:satellite_lae}
    
    In \autoref{sec:img-plane} we reported the discovery of a continuum counterpart to a local maximum of \lya~SB in the MUSE NB image labeled \arc-LAE. In this section we argue that \arc-LAE is a satellite of the main system composed of galaxies A and B, rather than another UV clump in the ISM of galaxy B. According to our best-fit lens model, this source has an average magnification of $\mu = 14$ and lies $\sim\SI{3}{\kilo\parsec}$ away from the UV centroid of galaxy B (see \autoref{fig:src_plane_maps}). This is approximately twice the exponential UV scale length of the continuum of galaxy B, further away than any of the UV clumps identifiable in that galaxy. Also, the candidate counterpart appears spatially resolved in the F606W image, with an exponential scale length of $250(14 / \mu)^{\frac{1}{2}}\,\si{\parsec}$, unlike the  clumps of galaxy B, which are all pointlike. \arc-LAE has an intrinsic \lya~luminosity of \SI{1.0+-0.2e41}{\erg\per\second}, representing the 2\% of the parent LAH luminosity. The F606W photometry implies an absolute UV magnitude of $M_{1500} \approx -16.7$, which lies at the faint end of the luminosity function of LAEs at this redshift \citep{Ouchi2008LyaLuminosityFunction, Kusakabe2020MuseHUDFLyaFraction}. In fact, the UV continuum at this luminosity is extremely difficult to detect, and only a handful of objects have been robustly detected in lensed fields  \citep{Claeyssens2022LlamasPaper1} or in deep stacks from the UDF \citep{Maseda2018UltraFaintELGs}. Although with very low significance, the UV slope of this source is steeper than the mean slope of galaxies A and B,  which is an expected property of strong LAEs due to their young ages \citep[e.g.,][]{Nakajima2012MetallicityAndSFRofLAEs, Hagen2014SedLAEsHetdexPilot} and low dust attenuation \citep[e.g.,][]{Ono2010StellarPopsLAEsLargeArea, Stark2010SpectroscopyOfLBGs, Kojima2017AbundanceRatiosToRedshift2}.
    
    Theoretical models and simulations predict the presence of several satellites populating the outer parts of LAHs, and some authors propose that they contribute a significant fraction of the \lya~SB at large radii ($r\gtrsim 0.25 R_\text{vir})$ \citep{MasRibas2017ExtendedLyaSatellites, Mitchell2021TracingSimulatedCGMLya}. It is thus plausible that \arc-LAE is indeed a satellite of the \arc~system, made detectable by the chance alignment of the lensing caustic boosting its flux above the background.
    
    Unfortunately, we did not detect additional lines in the MUSE spectrum of \arc-LAE due to strong contamination from the foreground galaxy light at $\lambda_\text{obs} \gtrsim \SI{4800}{\angstrom}$. Without a systemic redshift for this galaxy, interpreting the line profile becomes even more difficult. Nevertheless, the line profile shows a  broad red tail, suggesting that \lya~photons coming out of \arc-LAE are also scattered in the CGM.

\section{Conclusion}
We have analyzed the spatial and spectral properties of the diffuse \lya~emission associated with a pair of lensed LBGs at $z\approx 3$. The remarkable brightness and extension of this system, together with high-quality MUSE observations, allowed us to probe in detail its physical nature. 

The system is composed of two main-sequence galaxies (labeled A and B) of similar stellar mass ($\approx\SI{e10}{\msun}$) and SFRs ($\approx \SI{10}{\msun\per\year}$) that are separated by less than \SI{15}{\kilo\parsec} when projected in the source plane, and by \SI{145}{\kilo\meter\per\second} in the velocity space, suggesting an interacting pair. The galaxies are associated with a single LAH of $L_{\lya}=\SI{6.2+-1.3e42}{\erg\per\second}$ which we decomposed into two S\'ersic profiles in the source plane with the largest component having a circularized half-light radius of \SI{19.4}{\kilo\parsec}. Despite its apparent \lya~brightness, the whole system has a rest-frame \lya~EW of only \SI{17+-2.7}{\angstrom}. Globally, the \lya~line exhibits a redshifted peak with an asymmetric red tail, typical of CGM-scale outflows.
        
We found significant \SI{+-200}{\kilo\meter\per\second} spatial variations of the line FWHM and peak shift velocity across the halo. The lowest values of the FWHM and peak velocity shift were preferentially found on top of the central galaxies. We also recovered a  correlation between these two spectral properties, in line with recent results of resolved LAHs. 
        
We divided the source plane emission into 24 spatial bins and fitted them with radiative transfer models in isotropic galactic wind geometries. At an average expansion velocity of \SI{211+-3}{\kilo\meter\per\second} and standard deviation of \SI{62.5+-3}{\kilo\meter\per\second} we found tentative evidence of structured gradients along the minor axis, which suggests the outflow has a preferred direction. Also, the best-fit models imply that a typical  \lya~photon in the halo encounters a neutral column density of $\sim\SI{e20}{\per\centi\meter\squared}$ integrated along its path. The existence of the outflow is further confirmed by the presence of blueshifted, asymmetric absorption lines of low-ionization metal species  in the UV spectrum of the central galaxies. In particular, we measured the \ion{Al}{2} $\lambda 1670$ absorption central velocity at \SI{-197+-4}{\kilo\meter\per\second} and \SI{-142+-2}{\kilo\meter\per\second} in A and B, respectively, in broad agreement with the velocities inferred from the wind models. Finally, the recovered systemic redshifts for the different source plane regions show a complex structure that could be also explained by interaction processes between the galaxies.
        
We explored different mechanisms that could be producing the extended emission besides the resonant scattering of photons produced in the central galaxies. \lya~production \textit{in situ} by gravitational cooling is disfavored since we found no indication of infalling gas motion (e.g., a dominant blue peak) assuming an average IGM transmission. However, a major contribution of fluorescent radiation is allowed by energy budget arguments but it otherwise remains unconstrained due to uncertainties in the \lya~escape fraction and the clumpiness of the CGM. Upcoming H$\alpha$ observations with JWST will be key to establishing the contribution of extended fluorescent radiation in \name.

The boost in spatial resolution provided by the lensing effect allowed us to detect the continuum counterpart of a faint ($M_{1500} \approx -16.7$) satellite that is a strong LAE ($\text{EW}_0=\SI{104+-19}{\angstrom}$) and contributes 2\% of the total \lya~luminosity.  Moreover, this source is resolved with an exponential scale length of $\approx\SI{250}{\parsec}$. This is one of the few observational hints that such satellites do indeed exist, and contribute to the \lya~luminosity.

\begin{acknowledgements}
  We thank the anonymous referee for helpful suggestions that which have contributed to improving the quality of this manuscript. We also warmly thank Lucia Guaita for her insightful comments and discussion. This work was partly funded by Becas-ANID scholarship \#21221511. N.T. and S.L. acknowledge support from FONDECYT grant 1191232. M.A. acknowledges support from FONDECYT grant 1211951, CONICYT + PCI + INSTITUTO MAX PLANCK DE ASTRONOMIA MPG190030, CONICYT+PCI+REDES 190194, and ANID BASAL project FB210003. E.J.J. acknowledeges support from the FONDECYT Iniciaci\'on en Investigaci\'on 2020 Project 11200263. This paper is based on observations made with the ESO Telescopes at the La Silla Paranal Observatory under program ID 101.A-036. This research is also based on observations made with the NASA/ESA HST obtained from the Space Telescope Science Institute, which is operated by the Association of Universities for Research in Astronomy, Inc., under NASA contract NAS 5–26555. These observations are associated with programs GO-12368 and GO-15378. Support for analysis of the data from HST Program GO-15378 was provided through a grant from the STScI under NASA contract NAS5-265555.
\end{acknowledgements}

\appendix
\restartappendixnumbering 
    \section{Resolved EW}\label{sec:ew}
    Another important observable is the rest-frame \lya~ EW, $W_{\lya}$, since it has been shown to correlate very strongly with the \lya~escape fraction $\left(f_\text{esc}^{\lya}\right)$ of a galaxy \citep[e.g.,][]{Harikane2018SilverrushV, Sobral2019LyaEscapeSFR}, defined as the ratio between the observed and intrinsic \lya~luminosies. However, measuring the EW is subject to some complexities when the line profile is composed of both emission and absorption components \citep[e.g.,][]{Kornei2010LyaEmissionInLBGs, Erb2019LensedLyaHST}, as is the case of \name. Here, we calculate the net or total EW dividing the \lya~flux by the expected continuum level at $\lambda_\text{rest}=\SI{1215.67}{\angstrom}$, which was extrapolated from a power-law fit to the continuum at $\lambda_\text{rest}\gtrsim \SI{1270}{\angstrom}$. Using the image plane apertures for A and B, we extracted the full MUSE spectra and fitted a power law described by $f_\lambda \propto \lambda^\beta$ to all line-free channels at $\lambda_\text{rest}\gtrsim \SI{1270}{\angstrom}$. We used uniform priors on $\beta$ and the normalization factor. This resulted in slopes $\beta$ of \num{-1.18+-0.15} for galaxy A and \num{-0.79+-0.21} for galaxy B. We used these values to extrapolate the demagnified F606W magnitudes $(\lambda_\text{pivot} = \SI{5921}{\angstrom})$ to the redshift \lya~wavelength (\SI{4771}{\angstrom}) continuum, finding a total (A+B) intrinsic flux level of $f_{4771} = \SI{1.03+-0.16e-18}{\erg\per\second\per\centi\meter\squared\per\angstrom}$. Then, we computed the EW by dividing the total delensed \lya~flux of the LAH, $F_{\lya}=\SI{6.9+-0.2e-17}{\erg\per\second\per\centi\meter\squared}$ (see \autoref{sec:spatial}), by the underlying continuum level obtained above. After propagating all uncertainties, this ratio yields $W_{\lya} = (\num{66+-10})(1 + z)^{-1}\,\si{\angstrom} = \SI{17.0+-2.7}{\angstrom}$.  Such a value is typical of LBGs \citep[e.g.,][]{Steidel2003LBGsAtRedshift3} and, according to the empirically calibrated relation of \citet{Sobral2019LyaEscapeSFR}, it implies a global $f_\text{esc}^{\lya}=\num{0.082+-0.018}$.
    
    However, with the available data it is also possible to investigate spatial variations of the EW across the arcs. The distribution of the EW can inform us about the homogeneity of the intervening gas. For example if the EW is uniform across the source, it could mean that the \lya~signal is processed by an approximately homogeneous slab of gas. If the EW distribution is clumpy or has gradients, a more complex geometry can be in place. 
    
    \begin{figure}[!hbt]
        \centering
        \includegraphics[width=\columnwidth]{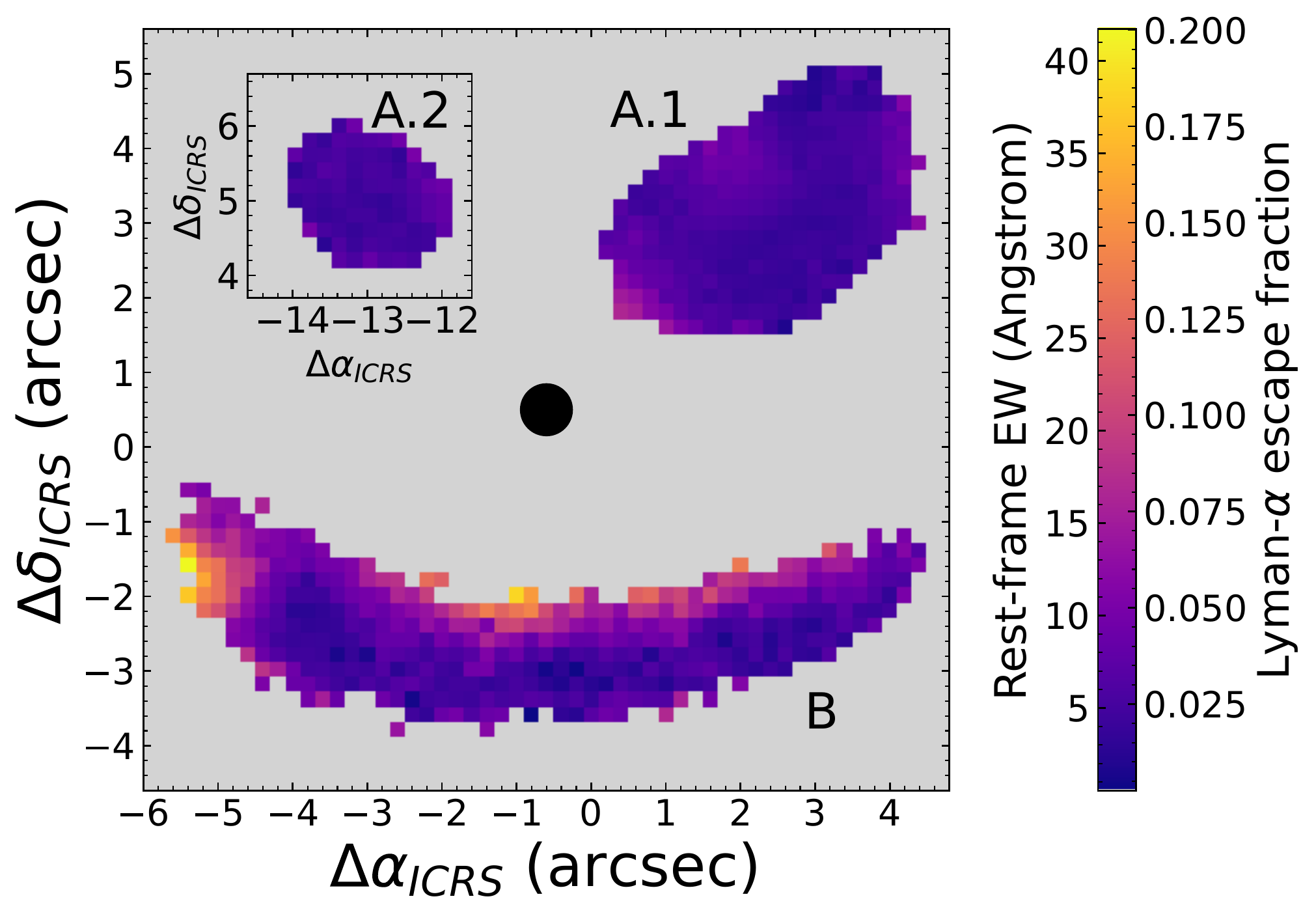}
        \caption{Image plane \lya~EW map displaying arcs A.1 and B and the counterimage A.2 (see inset panel). Spaxels without a continuum were masked out. The black circle in the middle of the panel indicates the FWHM size of the effective PSF. The right axis of the EW colorbar shows the corresponding \lya~escape fraction according to the \citet{Sobral2019LyaEscapeSFR} relation.}
        \label{fig:resolved_fesc}
    \end{figure}
    
    We constructed a map of the EW as the ratio between the \lya~NB and the extrapolated UV flux density at $\lambda_{\lya}$. To avoid discrepancies in the spatial resolution at different wavelengths, we estimated the UV slope in each spaxel by taking the ratio between PSF-homogenized images at $\lambda_\text{obs} = \SI{5300}{\angstrom}$ and $\lambda_\text{obs} = \SI{7060}{\angstrom}$ (rest-frame \SI{1350}{\angstrom} and \SI{1800}{\angstrom}, respectively) before extrapolating to $\lambda_{\lya}$. The images were obtained by averaging over the spectral axis on a $\approx\SI{200}{\angstrom}$ window centered at these two wavelengths and masking the channels with line absorption or emission. We created models of the MUSE-AO PSF by fitting a circular Moffat 2D profile to a single star near the center of the field and used them to convolve all images to a common PSF with FWHM=\SI{0.72}{\arcsecond}.  We computed the UV slope array from the ratio between the blue and red continuum images and then used it to extrapolate the continuum flux density to the \lya~wavelength. The EW map was finally constructed from the resulting continuum image at $\lambda_{\lya}$ and the NB image (see \autoref{fig:resolved_fesc}). This operation was restricted to the same continuum apertures defined in \autoref{sec:integrated_spectrum}.

    We observe that $W_{\lya}$ is mostly uniform toward galaxy A, with an average value of \SI{4}{\angstrom}, whereas galaxy B has values that range from \SI{3}{\angstrom} to \SI{42}{\angstrom}. The increased $W_{\lya}$ toward the northern and eastern edges of arc B can be explained by the large spatial offset between the continuum and the brightest knots of \lya~emission. As we mentioned above (\autoref{sec:img-plane}), the UV arc is offset by \ang{;;0.6} from the \lya~arc, and there is also a bright knot of \lya~emission to the east of the arc that has little overlap with the continuum. In \autoref{fig:resolved_fesc} we also provide the correspondence between EW and $f_\text{esc}^{\lya}$ as calibrated by \citet{Sobral2019LyaEscapeSFR}, but the interpretation of the EW map as an $f_\text{esc}^{\lya}$ map would require the relation to hold also for resolved regions.
    
    \section{AG Fitting}\label{sec:asym-gauss}
    
\begin{figure}[!hbt]
        \centering
        \includegraphics[width=\columnwidth]{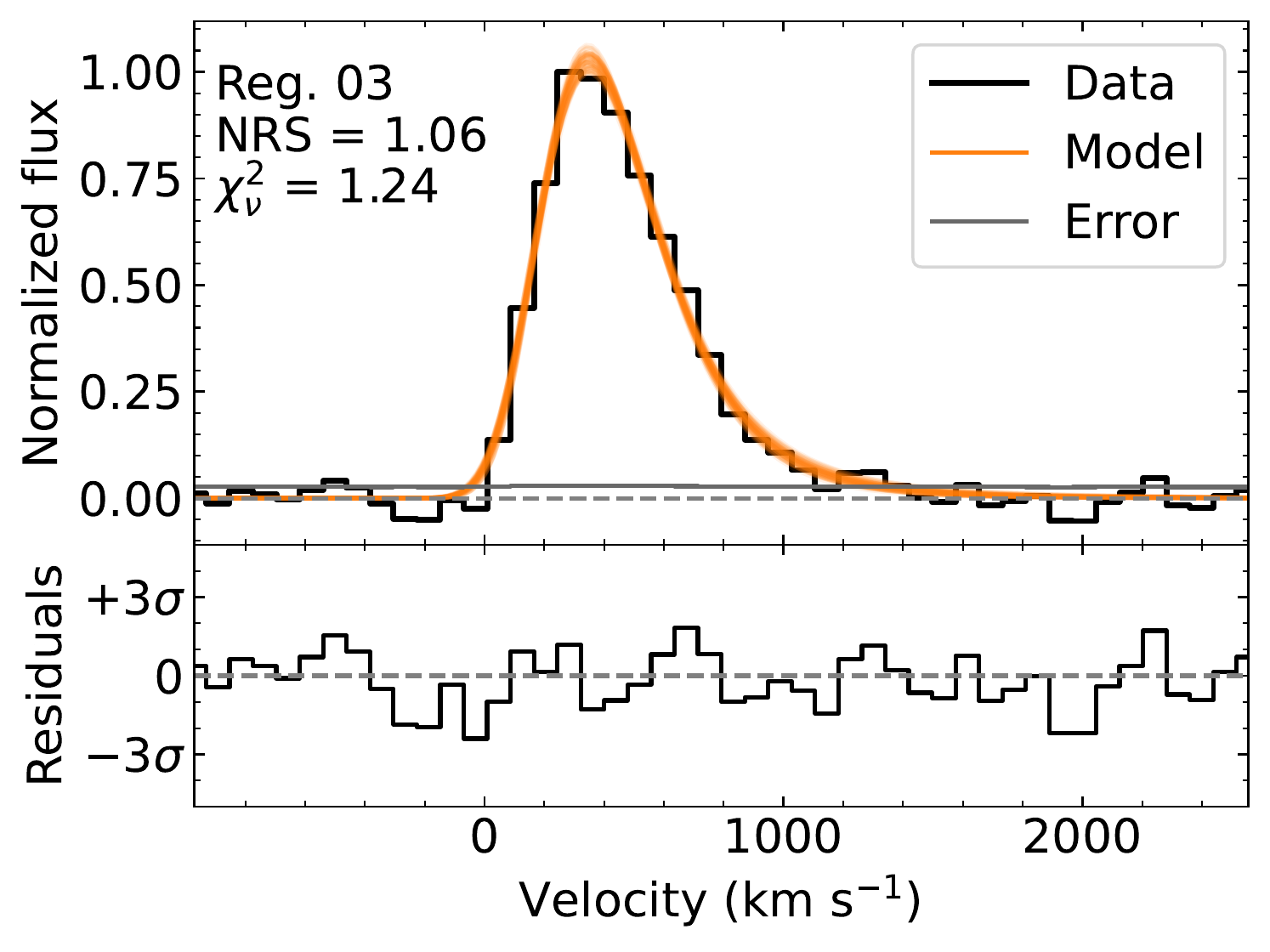}
        \caption{Example of AG fit to one of the extracted spectra. The solid orange curves show 50 random draws from the posterior probability distribution. The gray dashed line marks the zero-flux level. In the legend in the upper left corner, NRS stands for normalized residual scatter, which is the standard deviation of the residuals weighted by the error spectrum.}
        \label{fig:example_agfit}
    \end{figure}

    The AG profile as parameterized by \citet{Shibuya2014OriginStrongLya2} serves as a tool to measure basic properties of typical \lya~spectra, namely the amplitude, peak shift velocity, FWHM, and an asymmetry parameter that quantifies the skewness of the curve. The use of the AG profile is becoming increasingly common in the resolved LAH literature \citep{Claeyssens2019LyaSpectralVariations, Leclercq2020SpatiallyResolvedLyaHalosMHUDF, Claeyssens2022LlamasPaper1}. In this paper we fit \lya~spectra either from the integrated apertures (\autoref{sec:integrated_spectrum}), from individual resolved regions (\autoref{sec:spatial}) or from absorption lines (\autoref{sec:absorption}) using a standard MCMC posterior sampling scheme. For each fit, 
    we imposed a Gaussian likelihood for the residuals, which were computed from the difference between the data and the AG profile convolved with the LSF. We sampled the posterior distribution using \num{32} walkers and \num{2000} steps within the \verb|emcee| library \citep{ForemanMackey2013emcee} to obtain robust estimates of the parameter uncertainties and covariances. We set uniform priors on each of the five parameters (the four named above plus an amplitude parameter) and let them vary freely over their domain ranges. During optimization, we also kept track of the integrated flux of the model. The value for the peak shift velocity depends on the systemic redshift input by the user. An example of an AG fit is displayed in \autoref{fig:example_agfit}.
    
    \section{\lya~Profiles from Binned Regions}
    \begin{figure*}[!htb]
        \centering
        \includegraphics[width=0.9\linewidth]{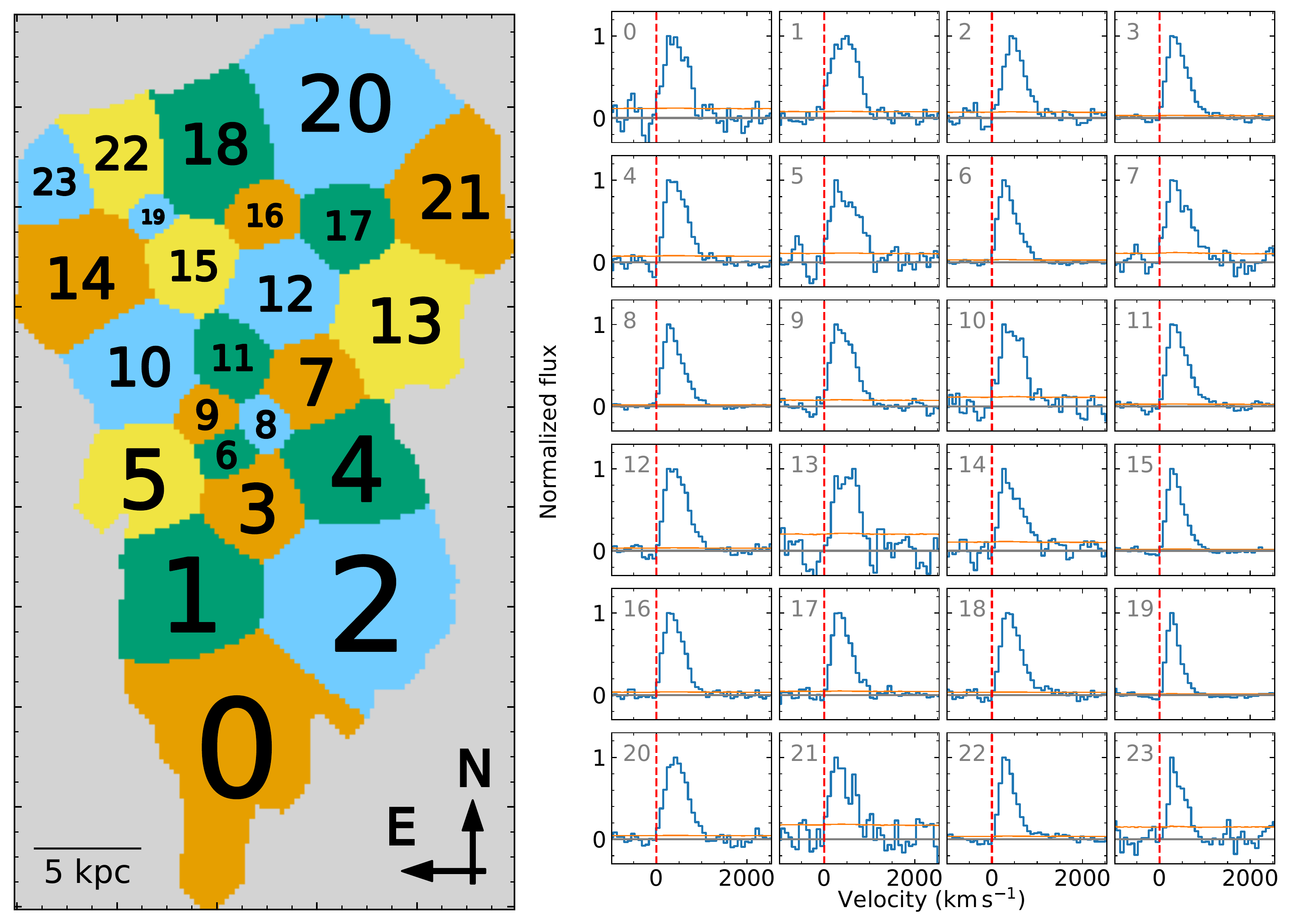}
        \caption{Left: 24-bin source plane tessellation with numbered bins. Right: Extracted spectra from the tessellated bins. The dashed red vertical line indicates the zero velocity at an arbitrary $z_\text{sys} = 2.9236$. The orange line shows the error spectrum associated with each bin.}
        \label{fig:profile_grid}
    \end{figure*}  
    
    \section{UV model residuals}\label{sec:uv-model}
    Here we present the residuals of the UV 2D model discussed in \autoref{sec:radial_profiles}.
    \begin{figure}
        \centering
        \includegraphics[width=\columnwidth]{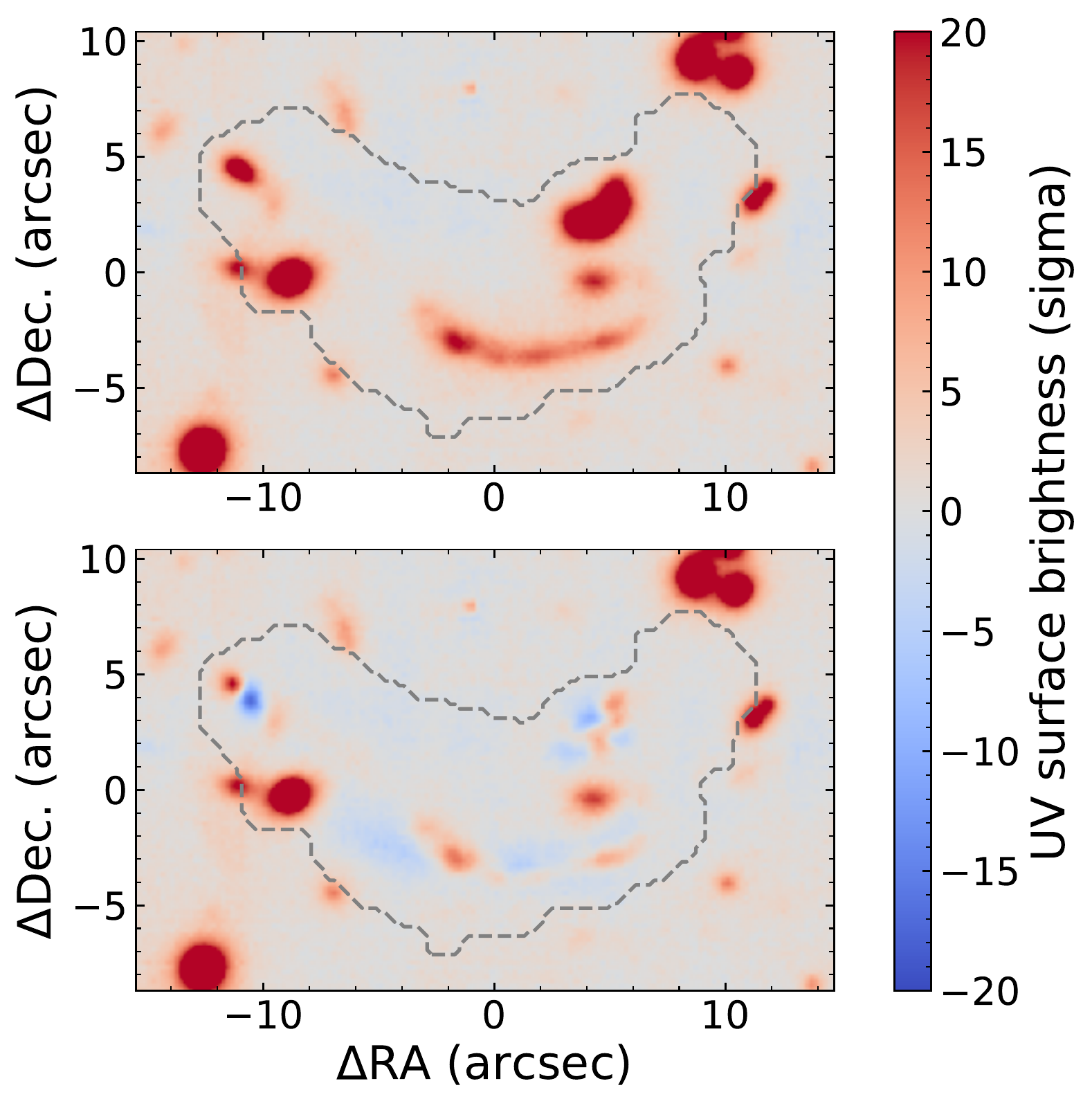}
        \caption{Upper panel: MUSE broadband image at $\lambda_\text{rest}\sim\SI{1600}{\angstrom}$. Lower panel: Same as above but with the best-fit model for galaxies A and B subtracted (after convolution with the MUSE PSF). In both panels the SB units are normalized to the background RMS and the dashed curve indicates the ``diffuse halo'' aperture (see \autoref{sec:integrated_spectrum}).}
        \label{fig:residual_uv}
    \end{figure}

\vspace{5mm}
\facilities{HST(ACS and WFC3), VLT:Yepun(MUSE)}
\software{Astropy \citep{Astropy2013PaperI, Astropy2018PaperII}, MPDAF \citep{Bacon2016MPDAF}, matplotlib \citep{Hunter2007MatplotlibPaper}, SciPy \citep{Virtanen2020SciPy}, emcee \citep{ForemanMackey2013emcee}, FLaREON \citep{GurungLopez2019FLaREON}, EsoRex \citep{ESO2015CPL}, GALFIT \citep{Peng2002Galfit1, Peng2010Galfit3}, BUDDI \citep{Johnston2017BUDDI}, vorbin \citep{Cappellari2003VorbinPaper}, scicm\footnote{\url{https://github.com/MBravoS/scicm}}}
\bibliography{main}{}
\bibliographystyle{aasjournal}

\end{document}